\documentstyle[aps,epsf]{revtex}
\renewcommand{\baselinestretch}{1.0}
{\def\theequation{A\arabic{equation}}}{}
{\def\theequation{B\arabic{equation}}}{}
\begin{document}
%
%
%
\title{
New properties of $z$-scaling:\\
flavor independence and saturation  at low $z$\\[10mm]
}
\author{ I.~Zborovsk\'{y}\footnote{Electronic address: zborovsky@ujf.cas.cz}}
\address{ Nuclear Physics Institute,
Academy of Sciences of the Czech Republic\\
250 68 \v {R}e\v {z}, Czech Republic}
\vskip 0.5cm
\author{M.V.~Tokarev\footnote{Electronic address: tokarev@sunhe.jinr.ru}}
\address{
Joint Institute for Nuclear Research\\
141980 Dubna, Russia}
\maketitle

\begin{abstract}
Experimental  data on inclusive cross sections of particles
produced in high energy proton-(anti)proton collisions at ISR,
RHIC, and Tevatron are analyzed in the framework of $z$-scaling.
New features of the scaling function $\psi(z)$ are established.
These are flavor independence of $\psi(z)$ including particles
with heavy flavor content and saturation at low $z$. The flavor
independence means that the shape of the scaling function
$\psi(z)$ is the same for different hadron species. The saturation
corresponds to flattening of $\psi(z)$ for $z<0.1$. Relations of
model parameters used in data $z$-presentation with some
thermodynamical quantities (entropy, specific heat, temperature)
are discussed. It is shown that behavior of the particle spectra at
low $z$ is controlled by a parameter $c$ interpreted as a specific
heat of the created medium associated with production of the
inclusive particle. The saturation regime of $\psi(z)$ observed at
low $z$ is assumed to be preferable in searching for phase
transitions of hadron matter and for study of non-perturbative QCD
in high energy proton-(anti)proton collisions at U70, RHIC,
Tevatron, and LHC.
\end{abstract}


\vskip 4.2cm

\begin{minipage}{14cm}
\hspace*{10.8cm} {\bf Preprint JINR}\\
\hspace*{11.cm}{\bf E2-2008-125}\\
\hspace*{11.0cm}{\bf Dubna, 2008}
\\[0.4cm]
\end{minipage}

\newpage

{\section{Introduction}}

The production of particles with high transverse momenta from
the collision of hadrons and nuclei at sufficiently high energies has
relevance to constituent interactions at small scales. In this regime,
it is interesting to search for new physical phenomena in elementary
processes such as quark compositeness \cite{Quarkcom}, extra
dimensions \cite{Extradim}, black holes \cite{Blackhole}, fractal
space-time \cite{Fracspace} etc.
Other aspects of high energy
interactions are connected with small momenta of secondary
particles and high multiplicities.
In this regime collective phenomena of particle production take place.
The search for new
physics in both regions is one of the main goals of investigations
at Relativistic Heavy Ion Collider (RHIC) at BNL and Large Hadron
Collider (LHC) at CERN \cite{RHIC_review}.
Experimental data on particle production at high energy and multiplicity
provide constraints for different theoretical models.
Processes with high transverse momenta of produced particles are
suitable for a precise test of perturbative Quantum Chromodynamics (QCD).
The soft regime is preferred for verification of non-perturbative QCD and
investigation of phase transitions in non-Abelian theories.

One of the methods allowing systematic analysis of data on
inclusive cross sections over a wide range of the collision energies,
multiplicity densities, transverse momenta, and angles of the
produced particles is based on the $z$-scaling observed in high
energy proton-(anti)proton collisions (see Ref.~\cite{Gen_Z} and
references therein). The approach to the description of the inclusive
spectra reflects the principles of locality, self-similarity, and
fractality of hadron interactions at a constituent level. It takes
into account the structure of the colliding objects, interaction
of their constituents, and processes of particle formation over a
wide scale range. The analyzed data include processes which take
place at nucleon scales as well as at small scales down to
$10^{-4}$~Fm. The presentation of experimental data in this
approach is given in terms of the scaling function $\psi(z)$ and
the scaling variable $z$. Both are constructed using the kinematical
variables, the experimentally measured inclusive cross section
$Ed^3\sigma/dp^3$, the multiplicity density $dN/d\eta$, and some model
parameters which allow physical interpretation.

It was shown\cite{Gen_Z} that the scaling behavior of $\psi(z)$ is
valid for different types of the produced hadrons. The scaling
function demonstrates two regimes. The first one is observed in the high-
and the second one in the low-$p_T$ region. The hard part of the inclusive spectra is
described by the power law, $\psi(z)\sim z^{-\beta}$, with a constant value of $\beta$ at large $z$.
The self-similar features of
particle production dictated by the $z$-scaling at large $z$ give
strong restriction on the asymptotic behavior of the
spectra in the high-$p_T$ region.
This provides suitable constraints
on phenomenology of parton distribution functions and
fragmentation functions which are needed to verify the
perturbative QCD in hadron collisions.
The soft regime of
particle production demonstrates flattening of the scaling
function $\psi(z)$ for small $z$.   The behavior of $\psi(z)$
at small $z$ is of interest to study the non-perturbative QCD
in the processes with low-$p_T$. We consider that this region is most
preferable in searching for phase transitions and study of
collective phenomena in multiple particle systems.

In this paper we show that in the high energy $pp$- and $p\bar p$-collisions
the shape of the scaling function at low $z$ is independent of the type
of the inclusive hadron including production of the hadrons with heavy
flavor content. A saturation of $\psi(z)$ with decreasing $z$ is
observed. The single parameter $c$ which controls the behavior of
$\psi(z)$ at low $z$ is interpreted as a specific heat of the
produced medium. The scaling in $pp$- and $p\bar p$-collisions is
consistent with a constant value of $c$. Search for a possible
change in this parameter is of interest especially
for soft processes with high multiplicities.
Such a change could be an indication of a phase transition in the matter produced
in high energy collisions of both hadrons and nuclei.

The paper is organized as follows.
A concept of the $z$-scaling and the method
of construction of the scaling function $\psi(z)$ are briefly described in Sec. II.
The properties of energy, angular, and multiplicity independence of the scaling
function are mentioned  in Sec. III.
The flavor independence and saturation of $\psi(z)$ at low $z$ are demonstrated in Sec. IV.
A microscopic picture of the constituent subprocess is analyzed in Sec. V.
A relation of the scaling variable $z$ to some thermodynamical quantities
is discussed in Sec. VI.
Here we consider possible manifestations of phase transitions
and their effects to the parameters used in the $z$-presentation of inclusive spectra.
Conclusions are summarized in Sec. VII.

\vskip 0.5cm
{\section{$z$-Scaling}}

In this paper we follow the version of the $z$-scaling presented in Ref. \cite{Gen_Z}.
Let us briefly remind  the basic ideas of this concept.
At sufficiently high energies, the collision of extended objects like hadrons
and nuclei is considered as an ensemble of individual interactions of their constituents.
The constituents are partons
in the parton model or quarks and gluons in the theory of QCD.
A single interaction of constituents is illustrated in Fig. 1.

\vskip 3.7cm
\begin{center}
\parbox{6cm}{\epsfxsize=4.8cm\epsfysize=4.8cm\epsfbox[95 95 400 400]
{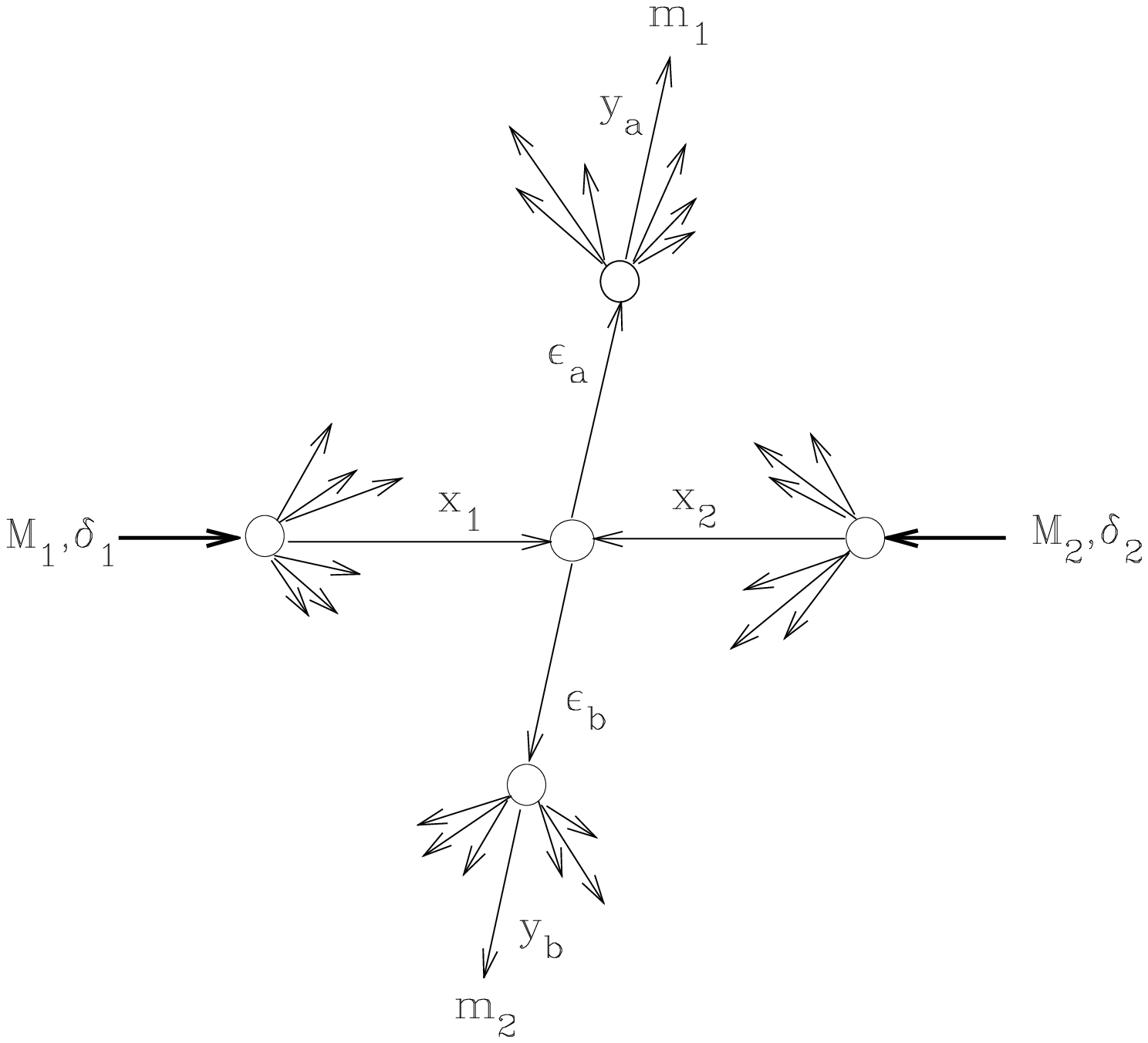}{}} \vskip -2.cm {\bf FIG. 1.} Diagram of the
constituent subprocess.
\end{center}

Structures of the colliding objects are characterized by
parameters $\delta_1$ and $\delta_2$.
The constituents of the incoming hadrons (or nuclei)
with masses $M_1, M_2$ and momenta $P_1, P_2$ carry their fractions $x_1, x_2$.
The inclusive particle
carries the momentum fraction $y_a$ of the scattered constituent
with a fragmentation characterized
by a parameter $\epsilon_a$.
A fragmentation of the recoil constituent is described
by $\epsilon_b$ and the momentum fraction $y_b$.
Multiple interactions are considered to be similar. This property
represents a self-similarity of the hadronic interactions
at the constituent level.

\vskip 0.5cm
{\subsection{Momentum fractions $x_1, x_2, y_a$, and $y_b$}}

The idea of the $z$-scaling is based on the assumption
\cite{Stavinsky} that gross features of an inclusive particle
distribution of the reaction
\begin{equation}
M_1 + M_2 \rightarrow m_1 + X
\label{eq:r1}
\end{equation}
can be described at high energies in terms of the kinematical
characteristics of the
corresponding constituent subprocess. We consider the
subprocess to be a binary collision
\begin{equation}
(x_{1}M_{1}) + (x_{2}M_{2}) \rightarrow (m_1/y_a) +
(x_{1}M_{1}+x_{2}M_{2} + m_2/y_b)
\label{eq:r2}
\end{equation}
of the constituents $(x_{1}M_{1})$ and $(x_{2}M_{2})$ resulting in
the scattered $(m_1/y_a)$ and recoil $(x_{1}M_{1}+x_{2}M_{2} +
m_2/y_b)$ objects in the final state.
The produced secondary objects transform into real particles after the constituent collisions.
The registered particle with the mass $m_1$ and the 4-momentum $p$
and its hadron counterpart,
moving in the opposite direction, carry the momentum fractions $y_a$ and $y_b$
of the scattered and recoil systems, respectively.
The momentum conservation law of the constituent subprocess
is connected with a recoil mass which we write in the form
\begin{equation}
(x_1P_1+x_2P_2-p/y_a)^2 =(x_1M_1+x_2M_2+m_2/y_b)^2.
\label{eq:r3}
\end{equation}
The associate production of $(m_2)$ ensures conservation
of the additive quantum numbers.
Equation (\ref{eq:r3}) is an expression of the locality of the hadron interaction
at a constituent level. It represents a kinematical constraint on the momentum
fractions $x_1$, $x_2$, $y_a$, and $y_b$ which determine a subprocess~(\ref{eq:r2}).

Structure of the colliding objects
and fragmentation of the systems formed in the scattered
and recoil directions are characterized by the parameters
$\delta_1,\delta_2$ and $\epsilon_a,\epsilon_b$, respectively.
We connect the structural parameters with the corresponding momentum fractions by
the function
\begin{equation}
\Omega(x_1,x_2,y_a,y_b)=
(1-x_1)^{\delta_1}(1-x_2)^{\delta_2}(1-y_a)^{\epsilon_a}(1-y_b)^{\epsilon_b}.
\label{eq:r4}
\end{equation}
Physical interpretation of $\Omega$ is given by its proportionality to
relative number of all such constituent configurations in the reaction (\ref{eq:r1})
which contain the configuration defined by the fractions $x_1,x_2,y_a,$ and $y_b$.
The function $\Omega$ plays the role of a relative volume which occupy these configurations
in the space of the momentum fractions.
It was found that the structural parameters $\delta_1$, $\delta_2$, $\epsilon_a$,
and $\epsilon_b$
have constant values at high energies.
They are interpreted as fractal dimensions in the corresponding space
of the momentum fractions.
For proton-proton collisions we set $\delta_1=\delta_2\equiv\delta$.
In the case of nucleus-nucleus collisions there are relations $\delta_1=A_1\delta$ and
$\delta_2=A_2\delta$, where $A_1$, $A_2$ are atomic numbers \cite{Z_A}.
We assume that the fragmentation of the objects moving in the scattered and recoil
directions can be described by the same parameter
$\epsilon_a=\epsilon_b\equiv\epsilon_F$ which depends
on the type $(F)$ of the inclusive particle.
For given values of $\delta$ and $\epsilon_F$,
we determine the fractions
$x_1$, $x_2$, $y_a$, and $y_b$ in a way to maximize the function
$\Omega(x_1,x_2,y_a,y_b)$, simultaneously fulfilling condition (\ref{eq:r3}).
The momentum fractions $x_1$ and $x_2$ obtained in this way can be
decomposed as follows
\begin{equation}
x_1=\lambda_1+\chi_1, \ \ \ \ \ \ \
x_2=\lambda_2+\chi_2,
\label{eq:r5}
\end{equation}
where $\lambda_{1,2}=\lambda_{1,2}(y_a,y_b)$ and $\chi_{1,2}=\chi_{1,2}(y_a,y_b)$ are
simple specific functions \cite{Gen_Z} of $y_a$ and $y_b$.
Using the decomposition, the subprocess (\ref{eq:r2}) can be
rewritten into a symbolic form
\begin{equation}
x_1+x_2\rightarrow (\lambda_1+\lambda_2) + (\chi_1+\chi_2).
\label{eq:r6}
\end{equation}
This relation means that the $\lambda$-parts of the interacting
constituents contribute to the production of the inclusive
particle, while the $\chi$-parts are responsible for the creation
of its recoil.
The maximum of the function (\ref{eq:r4}) with the condition (\ref{eq:r3})
can be obtained by searching for the
unconstrained maximum of the function
\begin{equation}
F(y_a,y_b)\equiv\Omega\left(x_1(y_a,y_b),x_2(y_a,y_b),y_a,y_b\right)
\label{eq:r7}
\end{equation}
of two independent variables $y_a$ and $y_b$.
There exists a single
maximum of $F(y_a,y_b)$ for every momentum $p$
in the allowable kinematical region.
Values of $y_a$ and $y_b$ corresponding
to this maximum have been determined numerically.
Having obtained the fractions $x_1$, $x_2$, $y_a$, and $y_b$,
we evaluate the function $\Omega$ according to (\ref{eq:r4}).
For fixed numbers $\delta$ and
$\epsilon_F$ we obtain in this way the maximal value of $\Omega$ for
every momentum $p$ of the inclusive particle.

Since the momentum fractions are determined by means of the maximization of the
expression (\ref{eq:r4}), they implicitly depend on
$\delta$ and $\epsilon_F$. The parameter $\epsilon_F$ enables
to take effectively into account also prompt
resonances out of which the inclusive particle of a given type may be created.
At fixed mass parameter $m_2$, larger values of $\epsilon_F$
correspond to smaller $y_a$ and $y_b$, which in turn give
larger ratios $m_2/y_b$ and $m_1/y_a$.
In our phenomenological approach this means that production of the
inclusive particle $(m_1)$ and its counterpart $(m_2)$ is a result of
fragmentation from larger masses which mimic in a sense processes with
prompt resonances.
Values of these parameters are determined in accordance with the experiment
and are discussed in the next sections.

\vskip 0.8cm
{\subsection{Scaling variable $z$ and scaling function $\psi(z)$}}

The self-similarity of hadron interactions reflects a property
that hadron constituents and their interactions are similar.
This is connected with dropping of
certain dimensional quantities out of the description of physical
phenomena. The self-similar solutions are constructed in terms of
the self-similarity parameters. We search for a solution
\begin{equation}
\psi(z) ={1\over{N\sigma_{inel}}}{d\sigma\over{dz}}
\label{eq:r8}
\end{equation}
depending on a single self-similarity variable $z$. Here
$\sigma_{inel}$ is an inelastic cross section of the reaction
(\ref{eq:r1}) and $N$ is an average particle multiplicity. The
variable $z$ depends on  momenta and masses of  the colliding and
inclusive particles, structural parameters of the interacting
objects, and dynamical characteristics of the produced system.
We define the variable $z$ as follows
\begin{equation}
z =z_0 \Omega^{-1},
\label{eq:r19}
\end{equation}
where
\begin{equation}
z_0 = \frac{  \sqrt {s_{\bot}}    }{(dN_{ch}/d\eta|_0)^c  m}
\label{eq:r10}
\end{equation}
and $\Omega$ given by (\ref{eq:r4}).
For a given reaction (\ref{eq:r1}), the variable $z$ is proportional to
the transverse kinetic energy $\sqrt {s_{\bot}}$ of the constituent subprocess
(\ref{eq:r2}) consumed on the production of the inclusive particle ($m_1$) and
its counterpart ($m_2$). The energy $\sqrt {s_{\bot}}$ is determined by the formula
\begin{equation}
\sqrt {s_{\bot}}=T_a + T_b,
\label{eq:r11}
\end{equation}
where
\begin{equation}
T_a=y_a(\sqrt { s_{\lambda}} -M_1\lambda_1-M_2\lambda_2)-m_1, \ \ \ \
T_b=y_b(\sqrt {s_{\chi}} -M_1\chi_1-M_2\chi_2)- m_2.
\label{eq:r12}
\end{equation}
The terms
\begin{equation}
\sqrt {s_{\lambda}} = [(\lambda_1P_1+\lambda_2P_2)^2]^{1/2},     \ \ \ \ \
\sqrt {s_{\chi}} = [(\chi_1P_1+\chi_2P_2)^2 ]^{1/2}
\label{eq:r13}
\end{equation}
represent the energy for production of the secondary objects moving in the scattered and
recoil direction, respectively.
The quantity $dN_{ch}/d\eta|_0$ is the corresponding multiplicity density of
charged particles in the central region of the reaction
(\ref{eq:r1}) at pseudorapidity $\eta=0$.
The multiplicity density in the central interaction region
is related to a state of the produced
medium. The parameter $c$ characterizes
properties of this medium \cite{Gen_Z}.
It is determined from multiplicity
dependence of inclusive spectra.
The mass constant $m$ is arbitrary and we fix it
at the value of nucleon mass.

The scaling function $\psi(z)$ is expressed in terms of the experimentally
measured inclusive cross section $Ed^3\sigma/dp^3$, the multiplicity
density $dN/d\eta$ at pseudorapidity $\eta$,
and $\sigma_{inel}$.
Exploiting the definition (\ref{eq:r8}) one can obtain the expression \cite{Gen_Z}
\begin{equation}
\psi(z) = -{ { \pi s} \over { (dN/d\eta) \sigma_{inel}} } J^{-1} E {
{d^3\sigma} \over {dp^3}},
\label{eq:r14}
\end{equation}
where $s$ is the square of
the center-of-mass energy and $J$ is the corresponding Jacobian.
The multiplicity density $dN/d\eta$ in the expression (\ref{eq:r14})
concerns particular hadrons species. It depends on the center-of-mass energy,
on various multiplicity selection criteria,
and also on the production angles at which the inclusive spectra were measured.
The procedure of obtaining the corresponding values of $dN/d\eta$ from the $p_T$ spectra
is described in  Ref. \cite{Gen_Z}.
The function $\psi(z)$ is normalized as follows
\begin{equation}
\int_{0}^{\infty} \psi(z) dz = 1.
\label{eq:r15}
\end{equation}
The above relation allows us to interpret the function $\psi(z)$ as a probability
density to produce an inclusive particle with the corresponding value
of the variable $z$.

\vskip 0.5cm
{\section{Properties of the scaling function}}

Let us remind  the properties of the $z$-presentation
of experimental data already found in proton-(anti)proton collisions
at high energies. These are the energy, angular, and multiplicity
independence of the scaling function $\psi(z)$ for different types
of hadrons, direct photons, and jets confirmed by numerous
data obtained at ISR, $\rm Sp\bar pS$, Tevatron, and RHIC.

\vskip 0.5cm
{\subsection{Energy independence of $\psi(z)$}}

The energy independence of the $z$-presentation of inclusive spectra means
that the shape of the scaling function
is independent on the collision energy $\sqrt s $ over a wide range
of the transverse momentum $p_T$ of the inclusive particle.
Results on the energy independence of the $z$-scaling for hadron production in
proton-proton collisions were presented in Ref. \cite{Gen_Z}.
The analyzed data \cite{FNAL1,ISR,ISR2,CDHW,FNAL2,STAR1,STAR3,STAR2,Adams}
include negative pions, kaons,
and antiprotons measured at FNAL, ISR, and RHIC energies.
The spectra were measured over a wide transverse
momentum range $p_{T}=0.1-10$~GeV/c.
The cross sections  decrease from $10^2$ to $10^{-10}$~mb/GeV$^2$ in this range.
The strong dependence of the spectra on the collision energy
increases with transverse momentum.
The independence of the scaling function
$\psi(z)$ on  $\sqrt{s}$ was found
for the constant values of the parameters $c=0.25$ and $\delta=0.5$ for all types
of the analyzed hadrons ($\pi, K, \bar p$).
The value of $\epsilon_F$ increases with the the mass of the produced hadron.
None of these parameters depends on the kinematical variables.
This was demonstrated for charged particles,
negative pions, kaons, and antiprotons in the range
of $\sqrt s = 19-200$~GeV.
The $z$-presentation of particle spectra produced in
proton-antiproton collisions was studied as well.
In this case the energy independence of the scaling function
was obtained for different set of parameters.
The scaling functions for proton-(anti)proton collisions
have similar shapes but differ in the region of large $z$.
The different values of the slope parameter $\beta$ ($\beta_{pp} > \beta_{p\bar p} $)
of the scaling function, $\psi(z)\sim z^{-\beta}$,
for charged hadron, direct photon, $\pi^0$-meson, and jet
production in $pp$ and $p\bar p$ collisions at high $z$ were found in Ref. \cite{Zflavor}.

\vskip 0.5cm
{\subsection{Angular independence of $\psi(z)$}}

The angular independence of the $z$-presentation of inclusive spectra means that
the shape of the scaling function is independent of the angle $\theta_{cms}$
of the produced particles over a wide range
of the transverse momentum $p_T$.
Results on the angular properties of the $z$-scaling in proton-proton collisions
are presented in Ref. \cite{Gen_Z}.
The analyzed experimental data \cite{CHLM} on angular dependence of inclusive spectra
include negative pions, kaons, and antiprotons measured at ISR energies.
The angles cover the range $\theta_{cms}=3^0-90^0$.
The central and fragmentation regions are distinguished by a different behavior
of differential cross sections.
The analysis included the transverse momentum spectra of charged hadrons
for $\theta_{cms}=5^0$  and $13^0$ \cite{BRAHMS}
obtained by the BRAHMS Collaboration at RHIC.
The data cover a wide range of the transverse momenta $p_T = 0.25-3.45$~GeV/c
of the produced hadrons at the collision energy $\sqrt s = 200$~GeV.
The angular independence of the scaling function $\psi(z)$
for charged particles, negative pions, kaons, and antiprotons was
obtained for the same values of the parameters $c$, $\delta$, and
$\epsilon_F$ which give the energy independence of $\psi(z)$.
This represents different values of $\epsilon_F$ for pions, kaons,...
and the same values of $\delta$ and $c$ for all of them.
The scaling function is
sensitive to the value of $m_2$ for small production angles $\theta_{cms}$.
This parameter was determined from the corresponding exclusive reactions
at the kinematical limit (for $x_1=x_2=y_a=y_b=1$).
Using Eq. (\ref{eq:r3}), this gives $m_2=m(\pi^+)$,
$m_2=m(K^+)$, and $m_2=m(p)$, for the inclusive production of
$\pi^-$, $K^-$, and antiprotons, respectively.
Note that the charged hadron multiplicity density  $dN_{ch}/d\eta|_{\eta=0}$
represents an angular independent factor in the definition of the variable $z$
which is the same for all particle species.
On the contrary, the scaling function  for pions, kaons, antiprotons,...
is normalized (\ref{eq:r14}) to the angular dependent multiplicity
density $dN/d\eta$ of the corresponding particles, respectively.

\vskip 0.5cm
{\subsection{Multiplicity independence of $\psi(z)$}}

The multiplicity independence of the $z$-presentation
of inclusive spectra means that
the shape of the scaling function $\psi(z)$ does not depend
on the multiplicity selection criteria characterized by the different
values of $dN_{ch}/d\eta$.
The multiplicity density
influences the shape of the inclusive spectra especially
at high transverse momenta \cite{Z_mult}.
At low $p_T$ the dependence of the cross sections on $dN_{ch}/d\eta$
concerns mainly the absolute values and much less the shape.
Results on the multiplicity independence of the $z$-scaling in proton-proton collisions
were presented in Ref. \cite{Gen_Z}.
The analysis was performed using the $K_S^0$-meson
and $\Lambda$-baryon spectra \cite{Witt}
obtained by the STAR Collaboration at $\sqrt s = 200$~GeV
for different multiplicity classes.
The charged multiplicity density was varied in the range $dN_{ch}/d\eta =1.3-9.0$.
The transverse momentum distributions were measured in
the central rapidity range $|\eta|<0.5$ up to the momentum $p_T = 4.5$~GeV/c.
The multiplicity independence of the scaling function provides a strong restriction on
the parameter $c$. Both data favor the same value of $c=0.25$ as was obtained
from the energy independence of the $z$-scaling in proton-proton collisions.
Similar applies to the $p_T$ distributions of charged particles \cite{Z_mult}
associated with the different multiplicity criteria.

An analysis of the spectra of direct photons and jets gives somewhat different results in the sense,
that the energy, angular, and multiplicity scaling was obtained by other
set of the parameters $c$, $\delta$, and $\epsilon_F$. Nevertheless, the corresponding
parameters do not depend on the kinematical variables, similarly as in the case
of the $z$-scaling for identified hadrons produced in proton-(anti)proton collisions.

\vskip 0.5cm
{\section{New properties of $z$-scaling  }

In this section we study possibility of a unified description of the
particle spectra of different hadrons using properties of their $z$-presentation.
The analysis is based on the observation that simultaneous energy, angular, and
multiplicity independence of the $z$-scaling for negative pions, kaons, and
antiprotons produced in proton-proton collisions gives the same shape
of the scaling function $\psi(z)$.
This flavor independence of $\psi(z)$ is confirmed here for other inclusive particles
including the particles with heavy quarks.
The independence in proton-antiproton interactions is observed with the exception of large $z$,
where the functions $\psi(z)$ for $pp$- and $p\bar p$-collisions mutually differ.
The transverse momentum spectra of $J/\psi$\cite{JPSI_B}
and $\Upsilon$\cite{UPS} mesons measured at the Tevatron
energies $\sqrt s = 1800$ and 1960~GeV
make it possible to investigate the behavior of the scaling function in the region
of very small $z$ (up to $10^{-3})$.
In this region we observe a saturation of $\psi(z)$ which can be approximated by a constant.
The saturation and flavor independence of the scaling function
for different hadrons with light and heavy quarks are
confirmation of the factorization
\begin{equation}
\frac{d^2\sigma}{dz d\eta}=\psi(z)\frac{d\sigma}{d\eta}
\label{eq:r16}
\end{equation}
of the inclusive cross sections.
This property of the $z$-scaling is valid for $pp$-collisions in a large range
of kinematical variables.
The factorization  of the differential cross section
in the variable $z$ and the pseudorapidity $\eta$ was demonstrated\cite{Gen_Z}
as the angular independence of the $z$-scaling for $\theta_{cms}=3^0-90^0$.
The scaling property at small angles gives strong restriction on the parameter $m_2$
which was found to be equal to the mass of the inclusive particle for the
negative pions, kaons, and antiprotons.
The same relation, $m_2=m_1$, is used here for all types of the inclusive hadrons.

\vskip 0.5cm
{\subsection{Flavor independence of $\psi (z)$}

Evidence of flavor independence of $z$-scaling was for the first time noted
in Ref. \cite{Zflavor}.
It was found that the value of the slope parameter $\beta$
of the scaling function
$\psi(z)$  at high $z$ is the same for different types
of produced hadrons ($\pi, K, \bar p$).
The hypothesis was later supported by the results
of an analysis of hadron ($\pi^{\pm,0},K,\bar{p}$) spectra
for high $p_T$ in $pp$- and $pA$-collisions.
Here we show that flavor independence of the $z$-presentation of hadron spectra
is valid for different hadrons over a wide range of the variable $z$.
Hence we exploit the scaling transformation
\begin{equation}
z \rightarrow \alpha_F z, \ \ \psi \rightarrow \alpha_F^{-1} \psi
\label{eq:r17}
\end{equation}
for comparison of the shape of the scaling function $\psi(z)$ for different hadron species.
The parameter $\alpha_F$ is a scale independent quantity.
The transformation does not change the shape of $\psi(z)$.
It preserves the normalization equation (\ref{eq:r15}) and
does not destroy the energy, angular, and multiplicity independence
of the $z$-presentation of particle spectra.

Figure 2(a) shows the
$z$-presentation of the spectra of negative pions, kaons,
antiprotons, and $\Lambda's$ produced in $pp$-collisions over the
range $\sqrt s = 19-200$~GeV and $\theta_{cms}=3^0-90^0$.
The symbols represent data on differential cross sections
measured in the central
\cite{FNAL1,ISR,ISR2,FNAL2,STAR3,STAR2,STAR5}
and fragmentation \cite{CHLM} regions, respectively.
The analysis comprises the inclusive spectra of particles \cite{BSM,BSM2} measured
up to very small transverse momenta ($p_T\simeq 45$~MeV/c for pions and
$p_T\simeq 120$~MeV/c for kaons or antiprotons).
One can see that the distributions of
different hadrons are sufficiently well described by a single curve over a wide
$z$-range ($0.01-30$).
The function $\psi(z)$ changes more than twelve orders of magnitude.
The solid lines represent the same curve shifted by multiplicative
factors for reasons of clarity.
The same holds for the corresponding data shown with the different symbols.
The indicated values of the parameter $\epsilon_F$
($\epsilon_{\pi}=0.2, \epsilon_{K}\simeq 0.3, \epsilon_{\bar p}\simeq 0.35,
\epsilon_{\Lambda}\simeq 0.4$) are consistent with the energy,
angular, and multiplicity independence of the $z$-presentation of spectra for different hadrons.
The parameters were found to be independent of kinematical
variables ($\sqrt s, p_T$, and $\theta_{cms}$).
The scale factors $\alpha_F$ are constants
which allow us to describe the $z$-presentation
for different hadron species by a single curve.
The estimated errors of  $\alpha_F$ are at the level of 20\%.

\vskip 2.5cm
\begin{center}
\hspace*{-0.5cm}
\parbox{6cm}{\epsfxsize=4.6cm\epsfysize=4.6cm\epsfbox[65 95 370 400]
{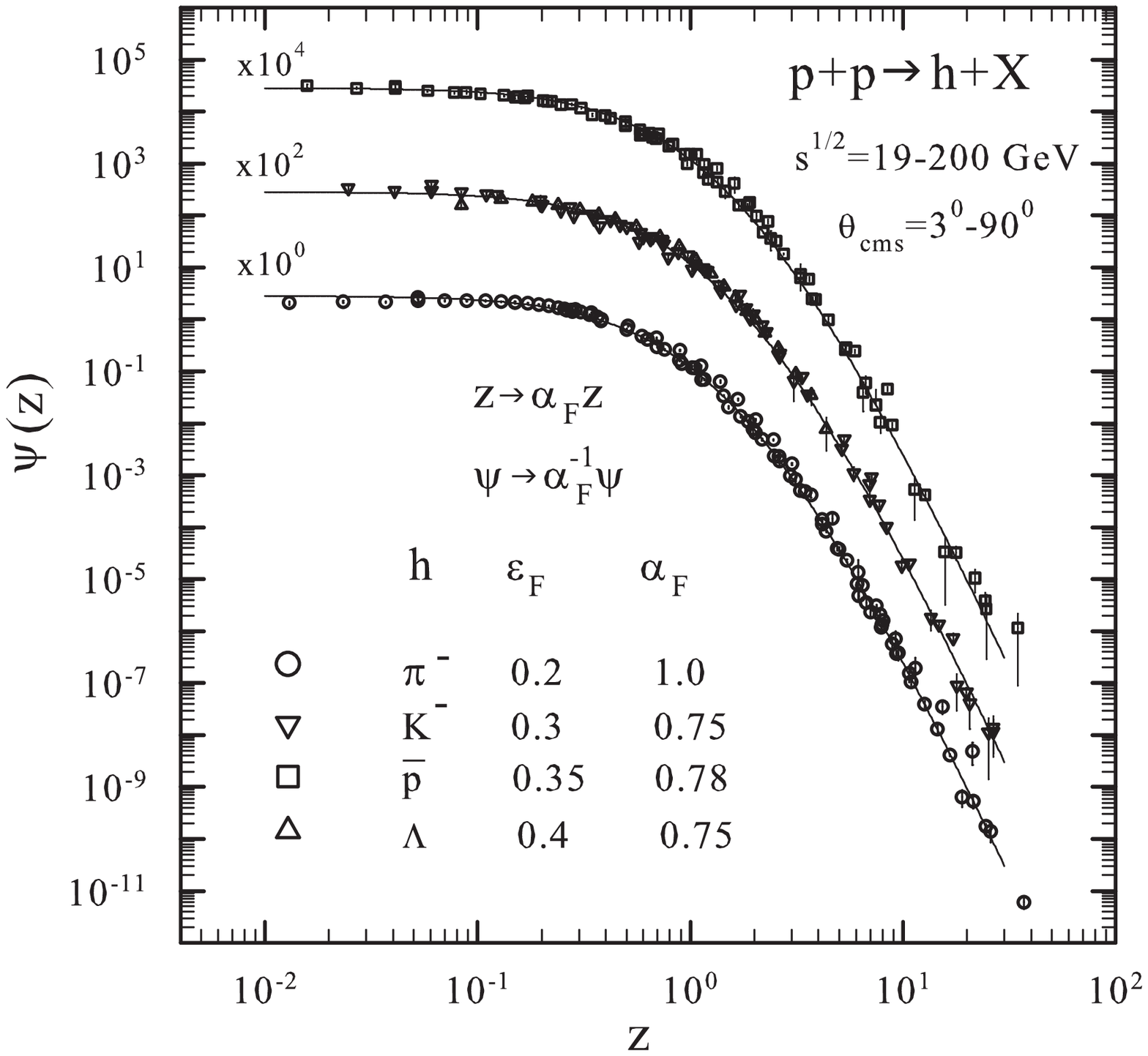}{}} \hspace*{2cm}
\parbox{6cm}{\epsfxsize=4.6cm\epsfysize=4.6cm\epsfbox[85 95 390 400]
{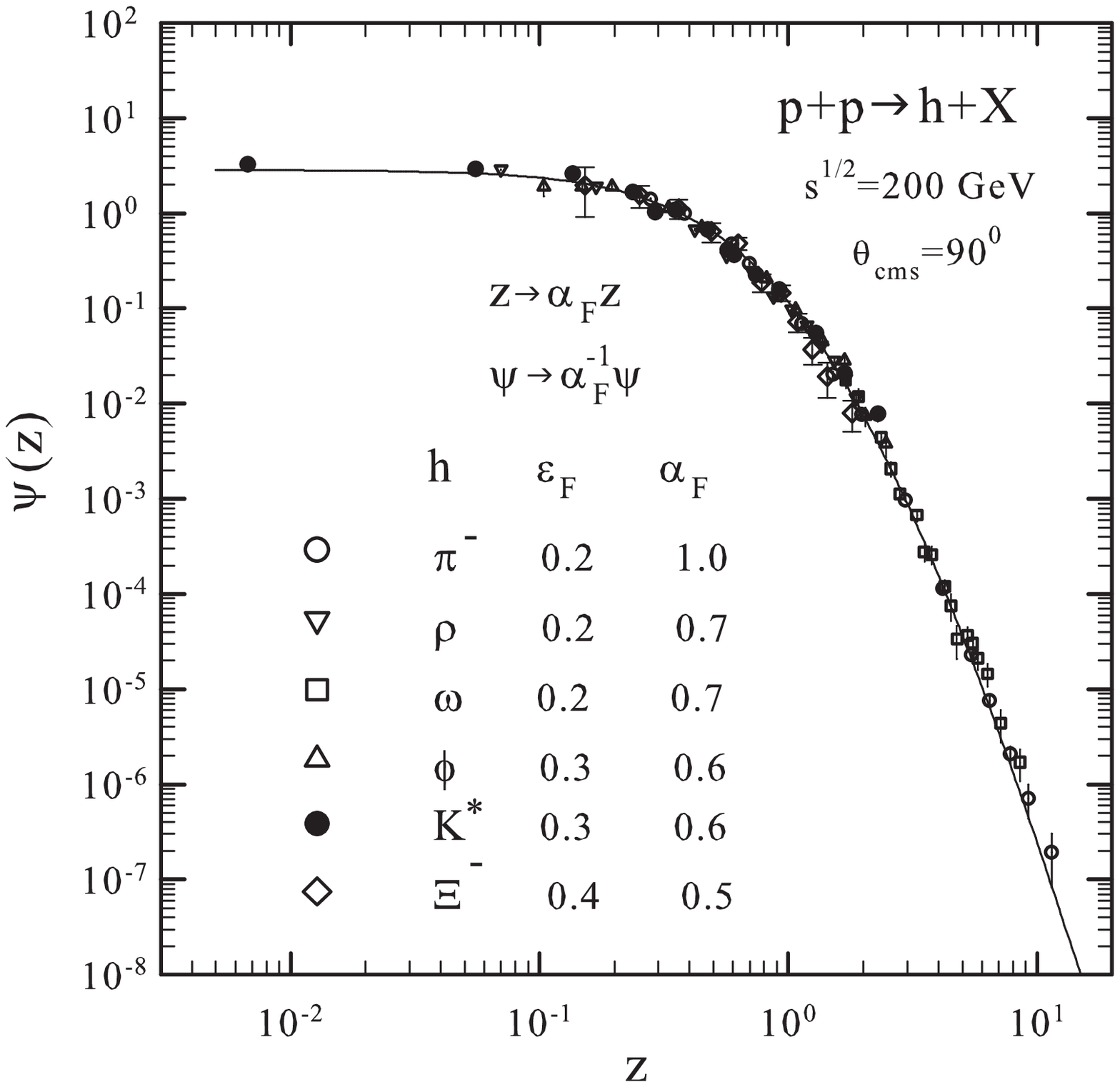}{}} \vskip -1.5cm
\end{center}
\vskip 0.4cm
\hspace*{36mm} (a) \hspace*{7.3cm} (b) \vspace*{8pt}\\
{\bf FIG. 2.} The flavor independence of $z$-scaling. The spectra
of $\pi^-$, $K^-$, $\bar p$, $\Lambda$ (a) and $\rho$, $\omega$,
$\phi$, $K^{*}$, $\Xi $ (b) hadrons produced in $pp$-collisions in
$z$-presentation. Data are taken from Refs.
\cite{FNAL1,ISR,ISR2,FNAL2,STAR3,STAR2,CHLM,STAR5,BSM,BSM2,RHO,OMEGA,PHI,K*,Xi}.
The solid line is obtained by fitting the data.
\vskip 1.0cm

Figure 2(b) shows similar results for other hadrons ($\rho, \omega, \phi, K^*, \Xi$)
produced in $pp$-collisions at $\sqrt s=200$~GeV and $\theta_{cms}=90^0$.
The experimental data \cite{RHO,OMEGA,PHI,K*,Xi} on inclusive spectra
are compared with the pion distributions \cite{STAR3} measured at RHIC.
The shape of $\psi(z)$ for these particles is described by the same curve (solid line) as
depicted in Fig. 2(a).
The black circle at the lowest $z \simeq 0.007$ corresponds to the STAR data \cite{K*}
on $K^*$ resonances
measured in the region where the scaling function is saturated.
Based on the obtained results we conclude
that RHIC data on $pp$-collisions confirm the flavor independence of the $z$-scaling including
the production of particles with very small $p_T$.

The inclusive spectra of heavier hadrons
($J/\psi, D^0,  B, \Upsilon$) \cite{JPSI_B,UPS,D0} obtained at the Tevatron
energies $\sqrt s=$1800 and 1960~GeV allow us to verify the new property of the $z$-scaling
in $p\bar p$-collisions.
The data include measurements up to small transverse momenta ($p_T\simeq125$~MeV/c for charmonia,
$p_T\simeq290$~MeV/c for bottomia, and $p_T\simeq500$~MeV/c for $B$-mesons).
Figure 3(a) shows the transverse momentum spectra of  $J/\psi, D^0,  B$,
and $\Upsilon $ mesons in the $z$-presentation. The scaling
function is the same for hadrons with light and heavy flavors
produced in $pp$- and $p\bar p$-collisions in the range $z = 0.001-4$.
This is indicated by the same line as in Fig. 2.
The corresponding values of the parameters $\alpha_F$ and $\epsilon_F$ are found to be
$\alpha_{J/\psi}=0.23$, $\alpha_{D^0}\simeq 0.23$, $\alpha_{B}\simeq 0.12$,
$\alpha_{\Upsilon (1S)} \simeq 0.15$ and
$\epsilon_{J/\psi}=1.$, $\epsilon_{D^0}\simeq 0.4$, $\epsilon_{B}\simeq 0.4$,
$\epsilon_{\Upsilon (1S)} \simeq 0.4$, respectively.
Figure 3(b) demonstrates results of combined analysis of the RHIC \cite{Jpsi_STAR}
and Tevatron \cite{JPSI_B} data
on $J/\psi$-meson spectra  measured in $pp$- and $p\bar p$-collisions
at different energies   $\sqrt s = 200, 1800, 1960$~GeV
and  angles $\theta_{cms}=22^0, 90^0$ in the $z$-presentation.
The solid line is the same one as shown in Figs. 2 and 3(a).
\\[0.5cm]
\vskip 1.5cm
\begin{center}
\hspace*{-0.5cm}
\parbox{6cm}{\epsfxsize=4.6cm\epsfysize=4.6cm\epsfbox[65 95 370 400]
{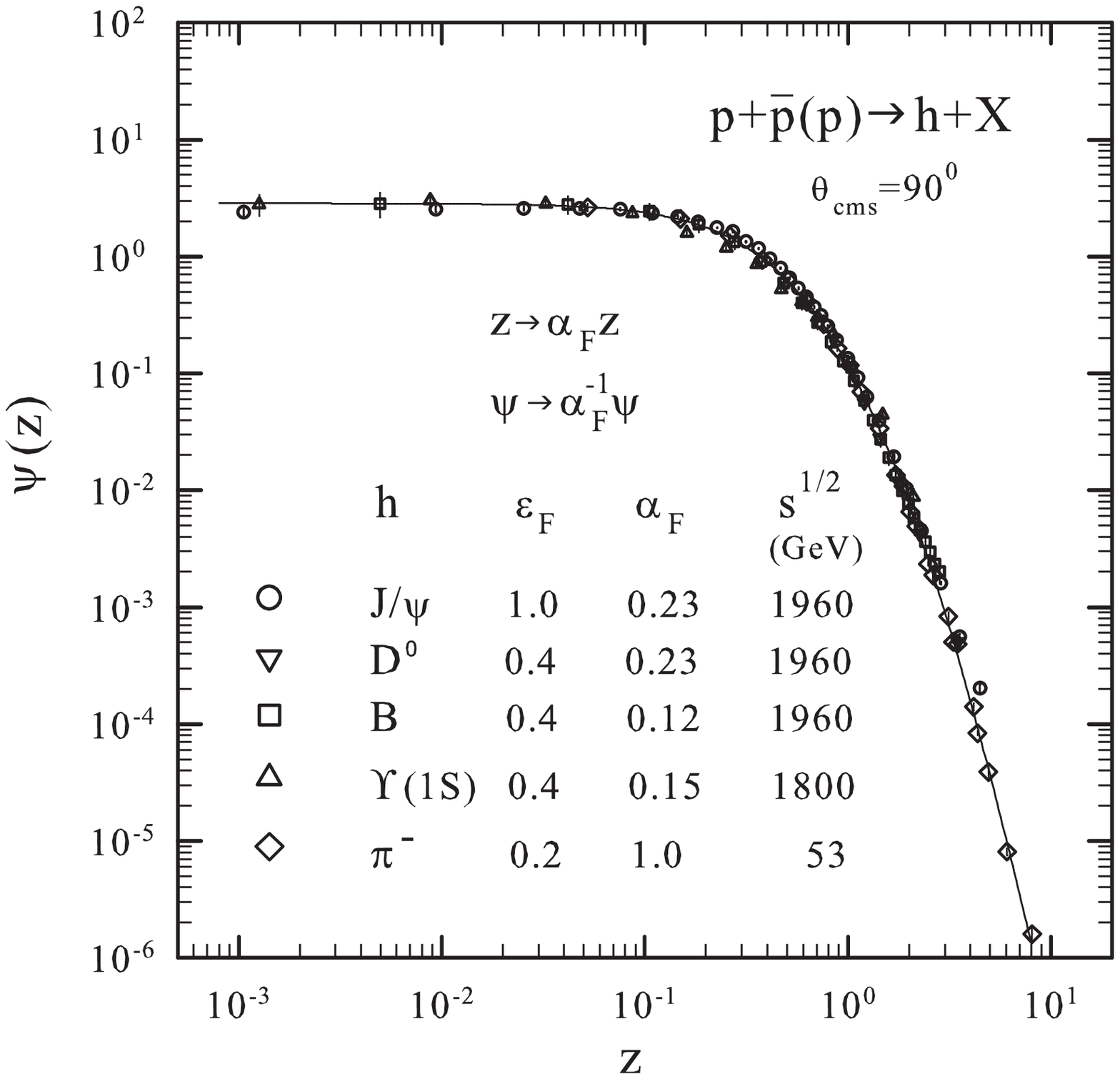}{}} \hspace*{2cm}
\parbox{6cm}{\epsfxsize=4.6cm\epsfysize=4.6cm\epsfbox[85 95 390 400]
{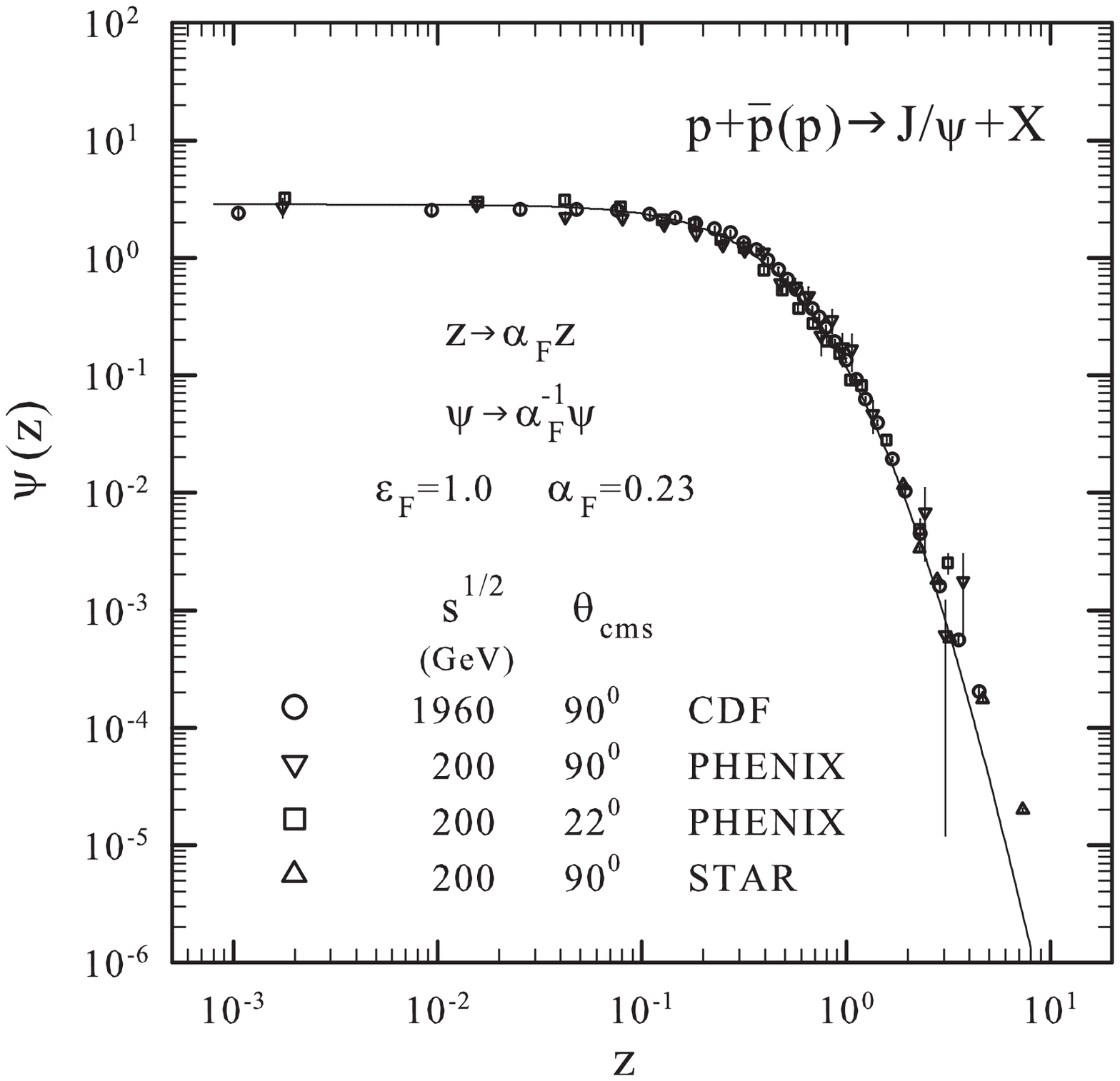}{}} \vskip -1.5cm
\end{center}
\vskip 0.4cm
\hspace*{36mm} (a) \hspace*{7.3cm} (b) \vspace*{8pt}\\
{\bf FIG. 3.}
The flavor independence of $z$-scaling.
The spectra of $J/\psi, D^0, B, \Upsilon$, $\pi^-$ (a)
and $J/\psi$ (b) mesons produced in $pp/p\bar p$-collisions
in $z$-presentation.
Data are taken from Refs. \cite{JPSI_B,UPS,D0,Jpsi_STAR}.
The solid line is the same as shown in Fig. 2.
\vskip 0.5cm

From the performed analysis  we conclude
that ISR, RHIC, and Tevatron data on inclusive spectra
manifest the flavor independence of the scaling function $\psi(z)$ over a wide range of $z$.
We would like to stress that the obtained result is based on
$p_T$ distributions of the cross sections $Ed^3\sigma/dp^3$
which reveal strong dependence on the energy, angle,
multiplicity, and type of the produced particle.

\vskip 0.5cm
{\subsection{Saturation of $\psi(z)$ }

The description of particle spectra in terms of the variable $z$ depends on the
parameters $\delta_{1,2}, \epsilon_F$, and $c$.
In the region of high $z$,
the production of different hadrons is characterized by different values of $\epsilon_F$.
A reliable estimation of this parameter requires processes with high $p_T$.
The same applies to $\delta_{1,2}$ which relates
to the structure of the colliding particles in the initial state.
On the contrary, the low-$z$ region corresponds to the small values of $p_T$ and low momentum fractions
$x_1$, $x_2$, $y_a$, and $y_b$.
In this region $z$ tends to zero as $p_T$ decreases
or $\sqrt s$ increases.
Here is $\Omega \simeq 1$ and the scaling variable can be approximated by
$z\sim s_{\bot}^{1/2}/({dN_{ch}/d\eta |_0})^c$.
The behavior of $\psi(z)$ at low $z$ is therefore governed by the single parameter $c$.

As seen from Figs. 2 and 3, the $z$-presentation of hadron distributions demonstrates
weak dependence on the variable $z$ in the soft (low $p_T$) region.
The data on the pion, kaon, antiproton and especially on $J/\psi$
and $\Upsilon$ spectra  manifest saturation
in the range $z=10^{-3}-10^{-1}$.
This regime is characterized by an approximate constant
behavior of the scaling function,
$\psi(z)\simeq const$.
The similar dependence is observed for other hadrons ($K^*, B, \rho$,..)
at low $p_T$.
A characteristic of the saturation is the slope $\beta$
of the scaling function which is diminishing with the decreasing $z$.
The value of $\beta $ is approximately zero for $z = 10^{-3}-10^{-1}$.

One can assume that the asymptotic behavior of $\psi(z)$ at $z\rightarrow 0$
is universal and reflects properties of the produced system consisting
of its constituents (hadrons or quark and gluons).
The universal scaling behavior in this region suggests
that mechanism of particle production at low $p_T$ is governed by
soft self-similar processes  which reveal some kind of a mutual equilibrium
leading to the observed saturation.

\vskip 0.5cm
{\section{A Microscopic Picture of Constituent Subprocesses}

The approach based on the $z$-scaling concept  allows us
to develop a microscopic
scenario of particle production in terms of the constituent interactions.
Here we discuss some features of this scenario.
We attribute the quantity $z$ to any inclusive particle
in the reaction (\ref{eq:r1}).
The scaling variable $z$ has character of a fractal measure.
It consists of the finite part
$z_0$ and of the divergent factor $\Omega^{-1}$.
The factor $\Omega^{-1}$ describes a resolution at which an
underlying subprocess
can be singled out of the inclusive reaction (\ref{eq:r1}).
The $\Omega(x_1,x_2,y_a,y_b)$ is proportional to number
of all configurations containing the incoming constituents
which carry the fractions $x_1$ and  $x_2$ of the momenta $P_1$
and $P_2$ and which fragment to the
inclusive particle $(m_1)$ and its counterpart $(m_2)$ with
the corresponding momentum fractions $y_a$ and $y_b$.
The parameters $\delta_1$ and $\delta_2$ have relation to fractal structure of
the colliding objects (hadrons or nuclei).
They are interpreted as fractal
dimensions of the colliding objects in the space of the momentum fractions.
The parameters $\epsilon_a$ and $\epsilon_b$ characterize the fractal behavior
of the fragmentation process in the final state.
A common property of fractal measures is their divergence with the
increasing resolution.
The scaling variable has this property
as the resolution $\Omega^{-1}$ goes to infinity,
\begin{equation}
z(\Omega) \rightarrow \infty \ ,\ \ \ \ \ \ \ {\rm if}  \ \ \ \ \
\Omega^{-1} \rightarrow \infty.
\label{eq:r18}
\end{equation}
It means that $z$ is the scale dependent quantity.
For an infinite resolution the whole reaction (\ref{eq:r1})
degenerates to a single subprocess (\ref{eq:r2}), all momentum
fractions become unity ($x_1=x_2=y_a=y_b=1$) and $\Omega=0$. This
kinematical limit corresponds to the fractal limit $z=\infty$. In
the general case, the momentum fractions are determined from a
principle of a minimal resolution of the fractal measure $z$. The
principle states that the resolution $\Omega^{-1}$ should be
minimal with respect to all binary subprocesses (\ref{eq:r2}) in
which the inclusive particle $m_1$ with the momentum $p$ can be
produced.  It fixes the values of the corresponding momentum
fractions $x_1, x_2, y_a$, and $y_b$ and singles out the most
effective binary subprocess which underlies the inclusive reaction (\ref{eq:r1}).

\vskip 0.5cm
{\subsection{Momentum fractions versus $p_T$ and $\sqrt s$}

The method of determination of the momentum fractions  makes it
possible to analyze  kinematics of the constituent interactions in
the framework of the developed approach. The study of $x_1-x_2$
and $y_a-y_b$ correlations and their dependencies on the collision
energy and transverse momentum of the inclusive particle gives us
possibility to look at microscopic picture of the underlaying
subprocesses.

The fractions $x_1$ and $x_2$ characterize amount of the energy (momentum)
of the interacting protons (antiprotons) carried by their constituents which
undergo the binary
collision (\ref{eq:r2}) that underlies the inclusive reaction (\ref{eq:r1}).
Figure 4(a) shows the dependence of the fraction $x_1$
on the transverse momentum $p_T$
of the negative pions, kaons, and antiprotons produced
in $pp$-collisions at $\sqrt s =$ 19, 53, 200~GeV
and $\theta_{cms}=90^0$.
The fractions $x_1$ and $x_2$  are equal each other in that case.
They increase nearly linearly with the transverse momentum $p_T$.
For fixed $p_T$, the fraction $x_1$ decreases as the collision energy $\sqrt s $ increases.
The $x_1$ is larger for the production of heavy particles as compared with light ones.
The kinematical limit of the reaction (\ref{eq:r1})
corresponds to $x_1=x_2=1$ at any collision energy
and for any type of the inclusive particle.
This can be seen in Fig. 4(a) for $\sqrt s =19$~GeV where the fraction $x_1$ approximates unity
at $p_T\simeq 9$~GeV/c for all three particles.

The correlation $x_1-x_2$ for the $\pi^-, K^-$, and $\bar p$ production
at the energy $\sqrt s =53$~GeV
and for various detection angles is shown in Fig. 4(b).
The central ($\theta_{cms}=90^0,58^0,40^0$)
and fragmentation ($\theta_{cms}=2.86^0$) regions
are distinguished by a different mutual behavior of $x_1$ and $x_2$.
The $x_1-x_2$ correlation at $\theta_{cms}=90^0$  is depicted as a strait line.
For other angles belonging to the central interaction region,
the correlation becomes more
complicated and depends on the particle type.
Both tendencies diminish with the increasing $p_T$  where the fraction
$x_2$ demonstrates linear dependence on $x_1$ over  a wide range.
This is a common feature  for all hadron species.
The situation is different in the fragmentation region.
The fraction $x_2$ is much less than $x_1$
and strongly depends on the type
of the inclusive particle at the small angle $\theta_{cms}=2.86^0$.
The heavier particle the larger $x_2$.
The increase of $x_2$ with $x_1$ is very slow in the range of small angles.
This is  changed dramatically near the kinematical limit ($x_1\simeq 1$)
where the fraction $x_2$ begins to grow larger.\\[0.5cm]
\vskip 1.5cm
\begin{center}
\hspace*{-0.5cm}
\parbox{6cm}{\epsfxsize=4.6cm\epsfysize=4.6cm\epsfbox[65 95 370 400]
{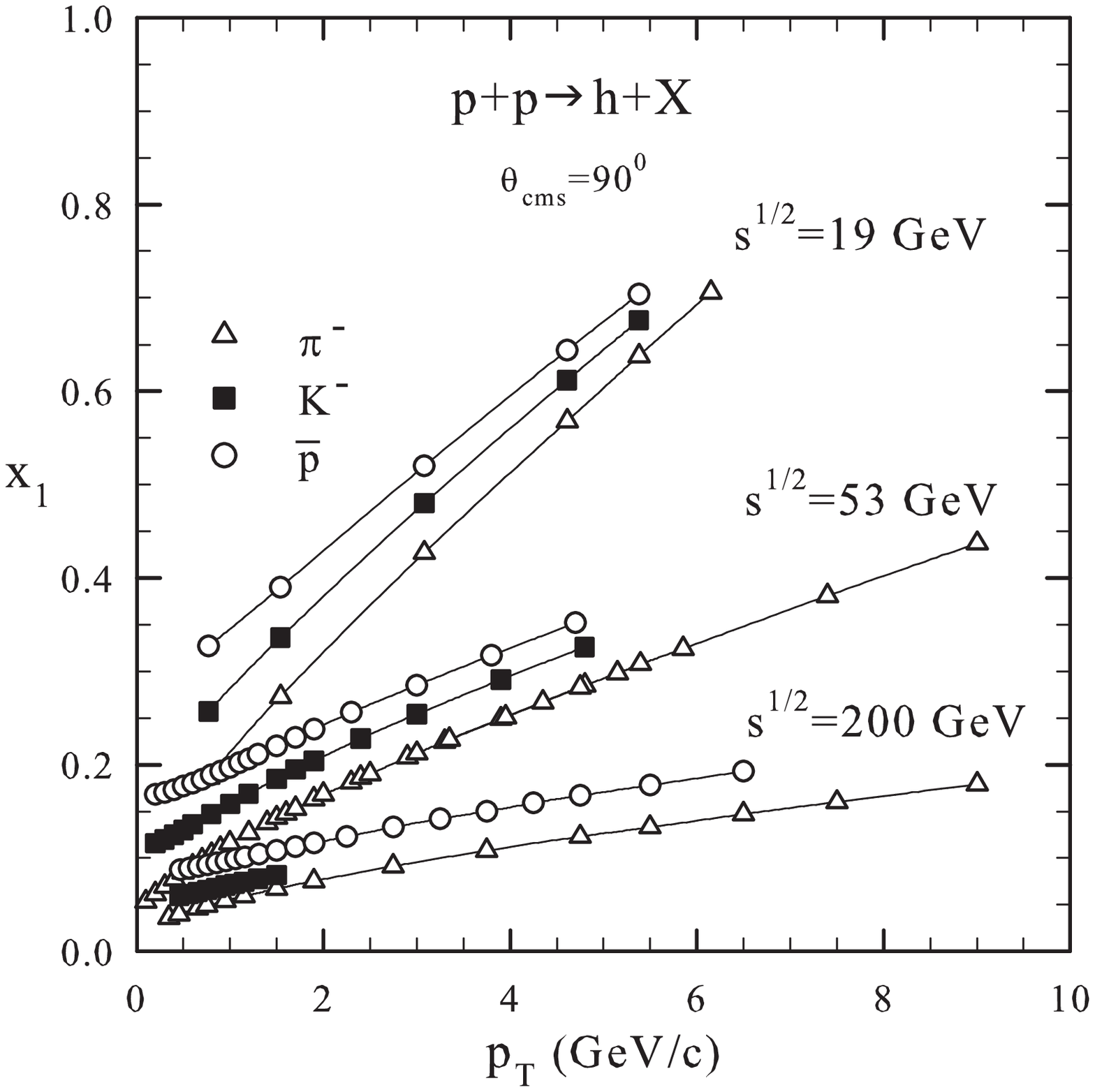}{}} \hspace*{2cm}
\parbox{6cm}{\epsfxsize=4.6cm\epsfysize=4.6cm\epsfbox[85 95 390 400]
{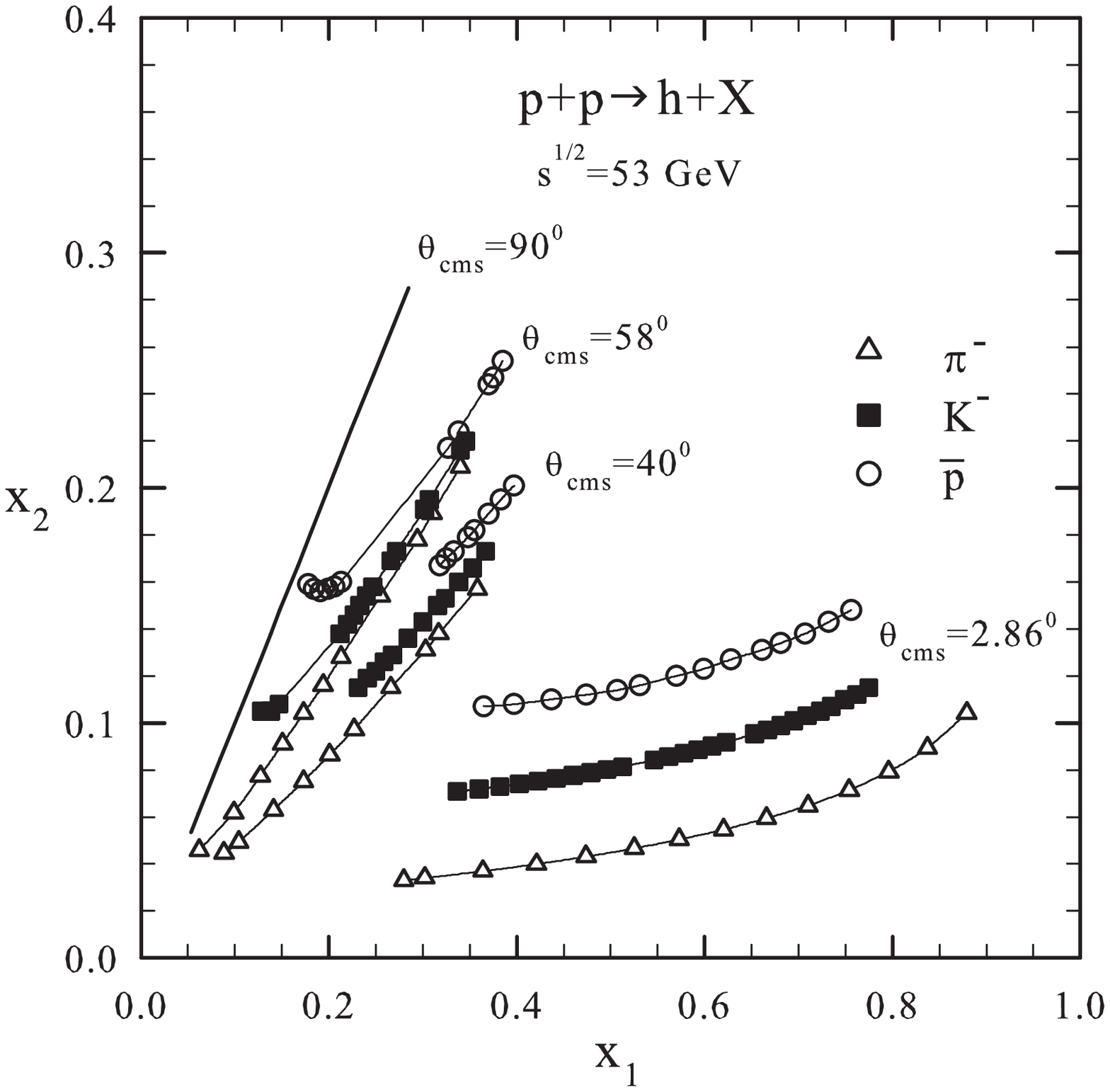}{}} \vskip -1.5cm
\end{center}
\vskip 0.4cm
\hspace*{36mm} (a) \hspace*{7.3cm} (b) \vspace*{8pt}\\
{\bf FIG. 4.}
(a) The dependence of the fraction $x_1$ on
the transverse momentum $p_T$ at $\theta_{cms}=90^0$. (b) The correlation between
the fractions $x_1$ and $x_2$ at $\sqrt s = 53$~GeV in
the central and fragmentation regions.
\vskip 0.5cm

The dependence of the momentum fractions $y_a$ and $y_b$ on the kinematical variables
($p_T, \theta_{cms}, \sqrt s $) describes features of the fragmentation process.
The fraction $y_a$ characterizes dissipation of the energy and momentum of the object
produced by the underlying constituent interaction into the near side of the inclusive particle.
This effectively includes energy losses of the
scattered secondary partons moving in the direction of the registered particle
as well as feed down processes from prompt resonances out of which the inclusive particle
may be created.
The fraction $y_b$ governs the recoil mass
in the constituent subprocess.
Its value characterizes the dissipation of the energy
and momentum in the away side direction of the inclusive particle.

\vskip 2.5cm
\begin{center}
\hspace*{-0.5cm}
\parbox{6cm}{\epsfxsize=4.6cm\epsfysize=4.6cm\epsfbox[65 95 370 400]
{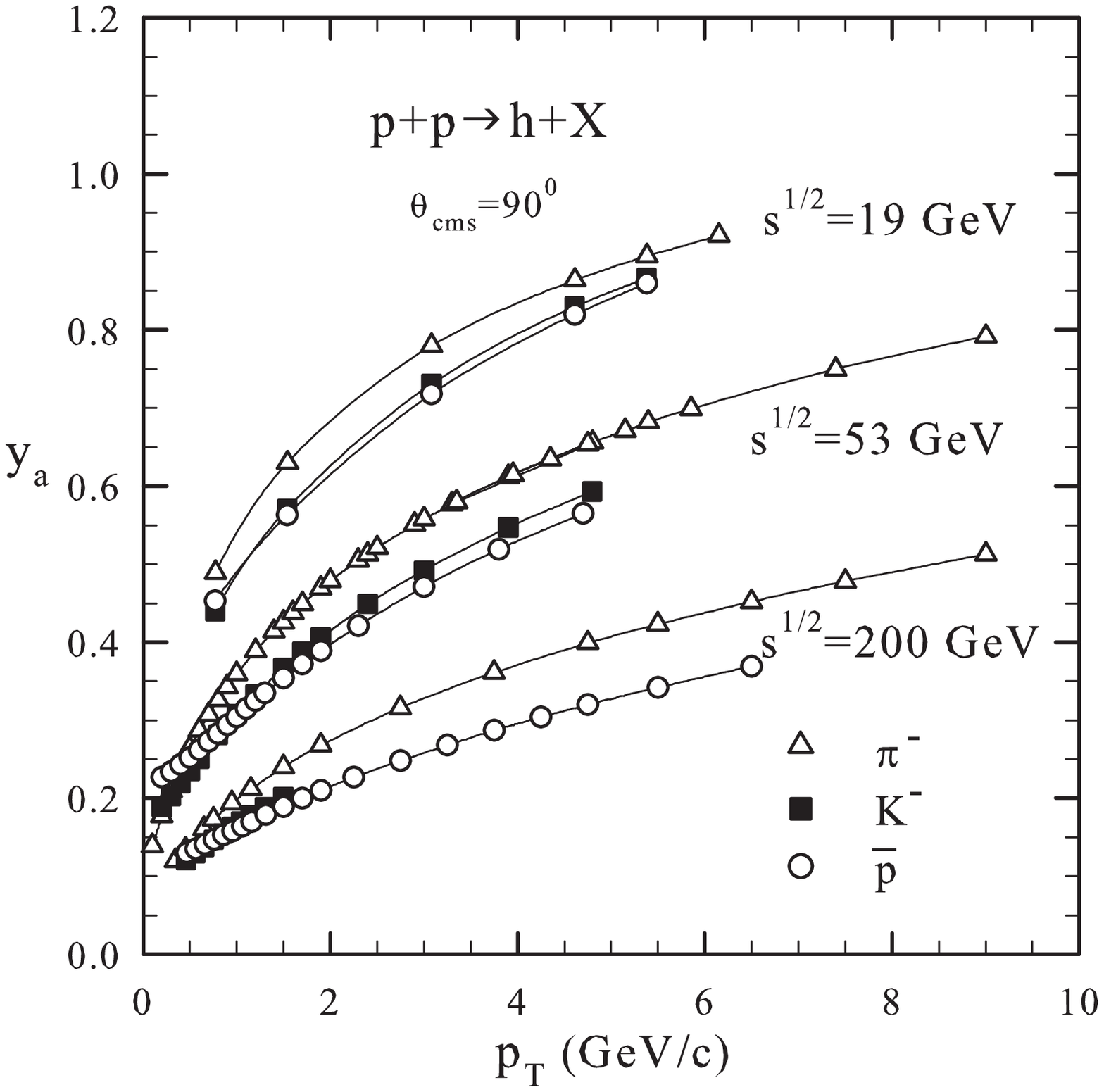}{}} \hspace*{2cm}
\parbox{6cm}{\epsfxsize=4.6cm\epsfysize=4.6cm\epsfbox[85 95 390 400]
{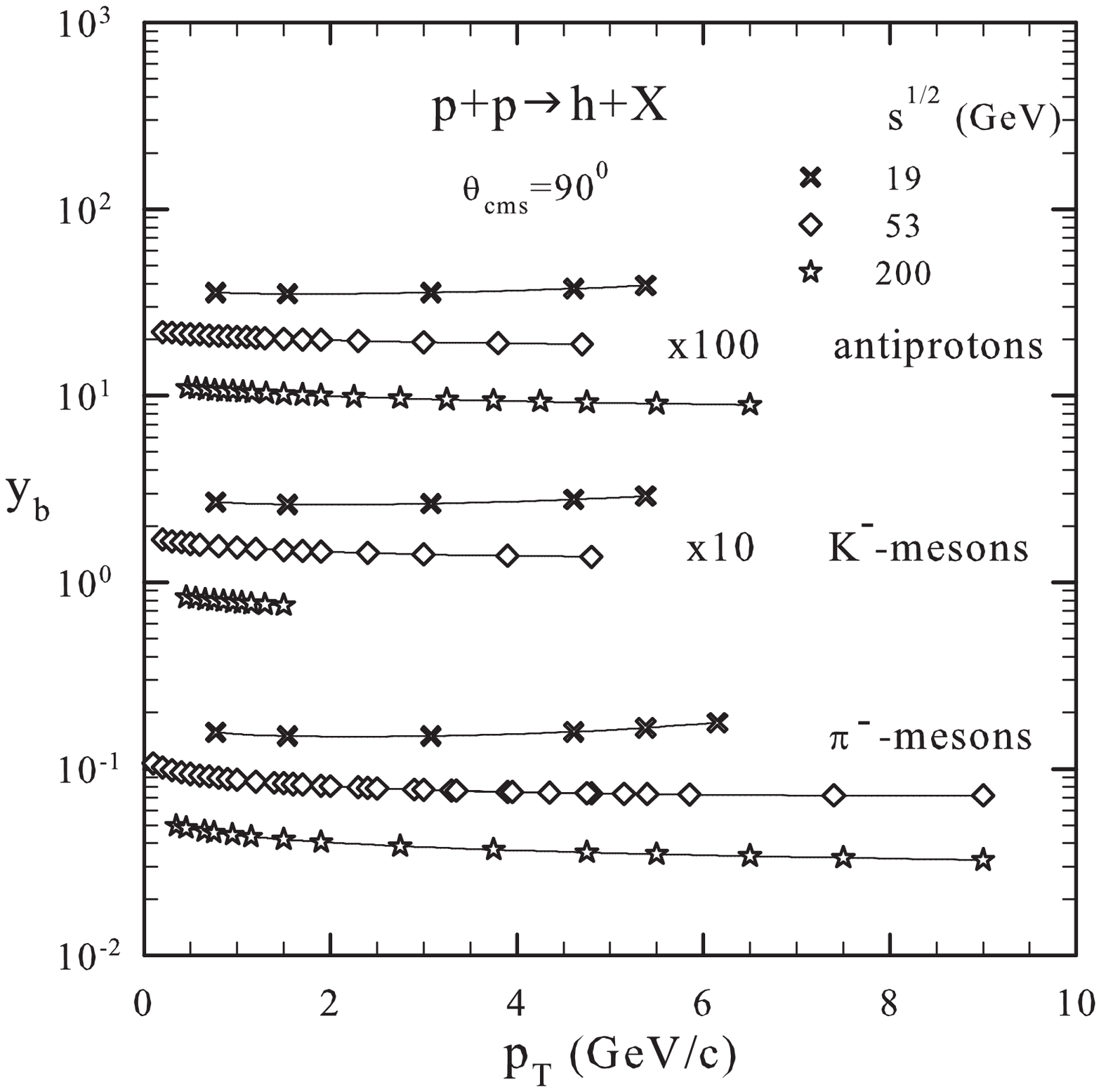}{}} \vskip -1.5cm
\end{center}
\vskip 0.4cm
\hspace*{36mm} (a) \hspace*{7.3cm} (b) \vspace*{8pt}\\
{\bf FIG. 5.}
The dependence of the fractions $y_a$ (a) and $y_b$ (b)
on the transverse momentum $p_T$
for  $\pi^-, K^-$, and $\bar p$ produced in the $pp$-collisions
at $\sqrt s = 19, 53$, and 200~GeV
in the central rapidity region.
\vskip 0.5cm

Figure 5(a) shows the dependence of $y_a$ on the transverse momentum $p_T$
for the negative pions, kaons, and antiprotons produced in $pp$-collisions
at the energy $\sqrt s =$ 19, 53, 200~GeV and $\theta_{cms}=90^0$.
All curves demonstrate a non-linear monotonic growth with $p_T$.
It means that the energy dissipation associated with the production
of a high $p_T$ particle
is smaller than for the inclusive processes with lower transverse momenta.
This feature is similar for all inclusive reactions at all energies.
Decrease of the fractions $y_a$
with the increasing collision energy is another property of the considered mechanism.
It corresponds to more energy dissipation at higher energies.
This can be due to the larger energy losses and/or due to the heavy prompt resonances.
The third characteristic is a slight decrease of $y_a$ with the mass of the inclusive particle.
It implies more energy dissipation for creation of heavier hadrons
as compared with hadrons with smaller masses.
The asymptotic value of $y_a=1$ is reached
at the kinematical limit for all particle species.

The dependence of $y_b$ on $p_T$ reflects kinematical properties of the recoil system.
One can see from Fig. 5(b) that $y_b$ is nearly independent of $p_T$.
It is smaller than $y_a$
and decreases  with the increasing collision energy $\sqrt s$.
The values of $y_b$ are larger for particles with higher masses.
The qualitative properties of the $p_T$ dependence of $y_b$
are similar for different hadrons.
The small values of $y_a$ mean that the momentum balance
in the production of an inclusive particle from a subprocess is more
likely compensated with many particles with smaller momenta than
by a single particle with a higher momentum moving in the opposite
direction.

\vskip 2.5cm
\begin{center}
\hspace*{-0.5cm}
\parbox{6cm}{\epsfxsize=4.6cm\epsfysize=4.6cm\epsfbox[65 95 370 400]
{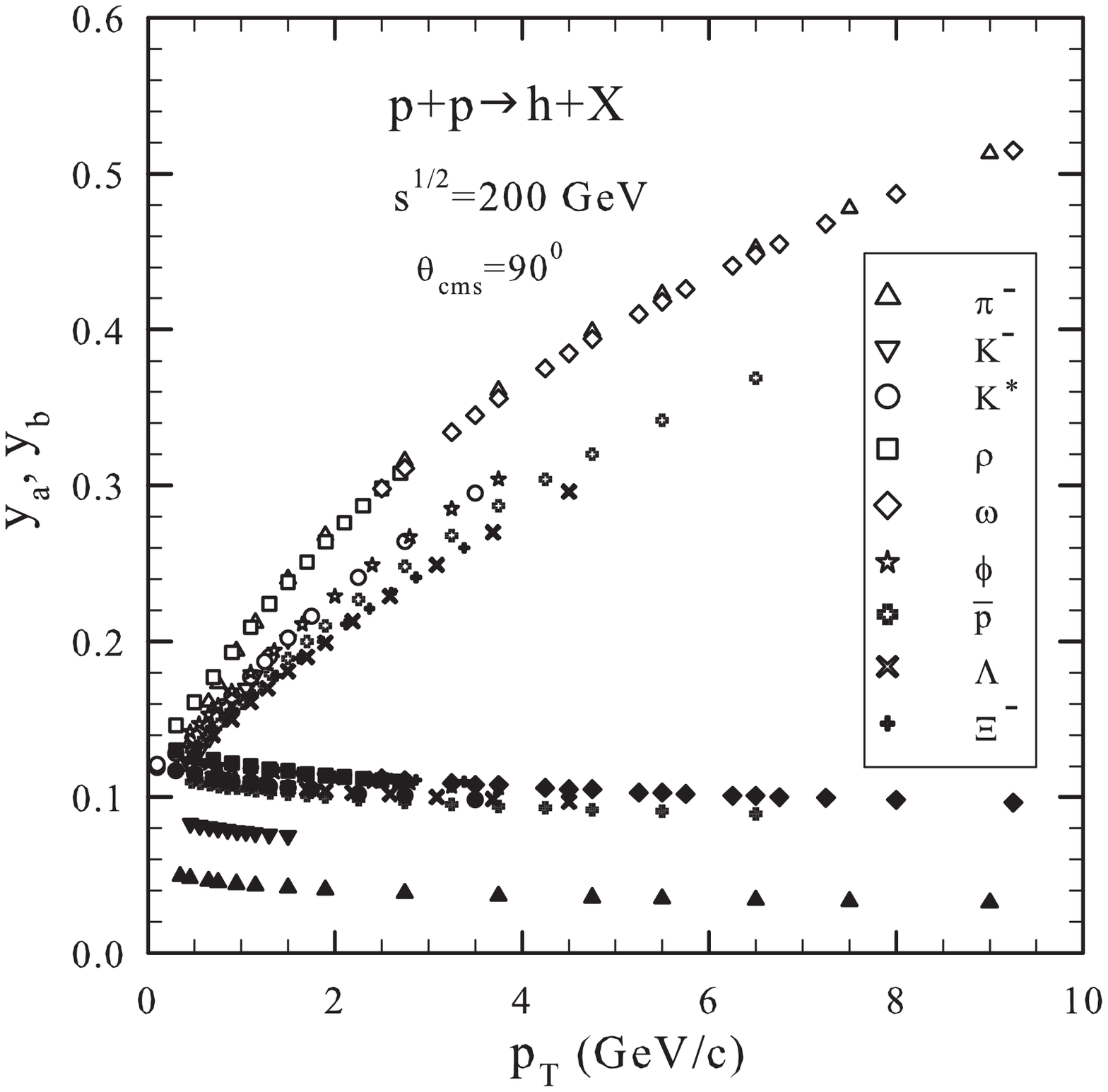}{}} \hspace*{2cm}
\parbox{6cm}{\epsfxsize=4.6cm\epsfysize=4.6cm\epsfbox[85 95 390 400]
{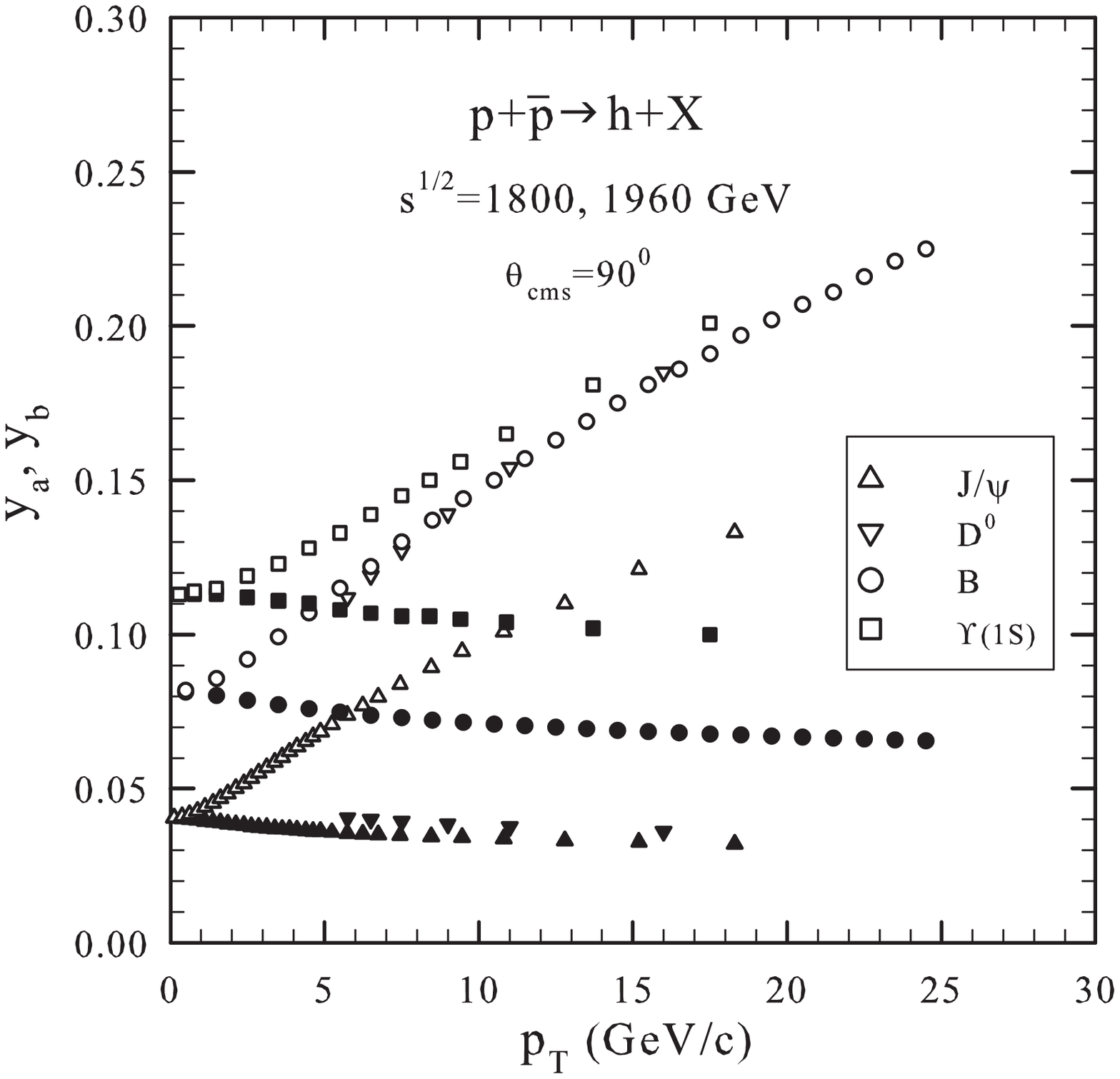}{}} \vskip -1.5cm
\end{center}
\vskip 0.4cm
\hspace*{36mm} (a) \hspace*{7.3cm} (b) \vspace*{8pt}\\
{\bf FIG. 6.}
The dependence of the  fractions $y_a$ and $y_b$  on the transverse momentum $p_T$
for (a) $\pi^-$, $K^-$, $K^*$, $\rho$, $\omega$, $\phi$, $\bar p$, $\Lambda$, $\Xi^-$ hadrons
produced in $pp$-collisions
at $\sqrt s = $~200~GeV  and  (b) $J/\psi$, $D^0$, $B$, $\Upsilon (1S)$ mesons
produced in $\bar pp$-collisions at $\sqrt s = 1800, 1960$~GeV.
\vskip 0.5cm

The $p_T$-dependence of $y_a$ and $y_b$ for other hadrons produced
in $pp$-collisions have similar behavior.
This is demonstrated in Fig. 6(a) on the data \cite{STAR3,STAR2,STAR5,RHO,OMEGA,PHI,K*,Xi}
at $\sqrt s =200$~GeV and $\theta_{cms}=90^0$ obtained at RHIC.
The hollow symbols demonstrate growth of $y_a$ with $p_T$.
The full symbols show flattening of $y_b$ in the range of $p_T=0.2-10$~GeV/c.
Note that both fractions become equal each other, $y_a\simeq y_b$,
for heavier particles and low transverse momenta.
This together with $m_1=m_2$ means that,
at low $p_T$, the objects produced in the constituent collision
into the near- and away-side direction have equal masses.
The feature is even more pronounced for particles with heavy quarks.
It is well seen from Fig. 6(b) where the momentum dependence of  $y_a$ and $y_b$ is demonstrated
for $J/\psi, D^0, B$, and $\Upsilon$-mesons produced
in $p\bar p$-collisions at the Tevaron energies  $\sqrt s = 1800$ and 1960~GeV.
The spectra correspond to the central rapidity range.
The $p_T$-dependence of $y_a$ and $y_b$ has similar behavior
as for data obtained at the RHIC energy.\\\\

There is, however, a particularity in the absolute values of the momentum fractions
which concern the $J/\psi$-meson production.
The corresponding curves for $y_a$ and $y_b$ lay much
below than what one could expect when compared with other particles, even at high $p_T$.
This is a consequence of the relatively large value of $\epsilon_{J/\psi}=1$ which
follows from the requirement of the energy independence of $\psi(z)$ for the charmonium production
(see Fig. 3).
In our approach it means that $J/\psi$-meson is produced from
larger masses accompanied with extra large dissipation of energy in the final state.
This exceptional property predestinate the $J/\psi$-meson to be a suitable probe in
$AA$-collisions where the energy losses can be sensitive to different phases of
the created matter.

\vskip 2.5cm
\begin{center}
\hspace*{-0.5cm}
\parbox{6cm}{\epsfxsize=4.6cm\epsfysize=4.6cm\epsfbox[65 95 370 400]
{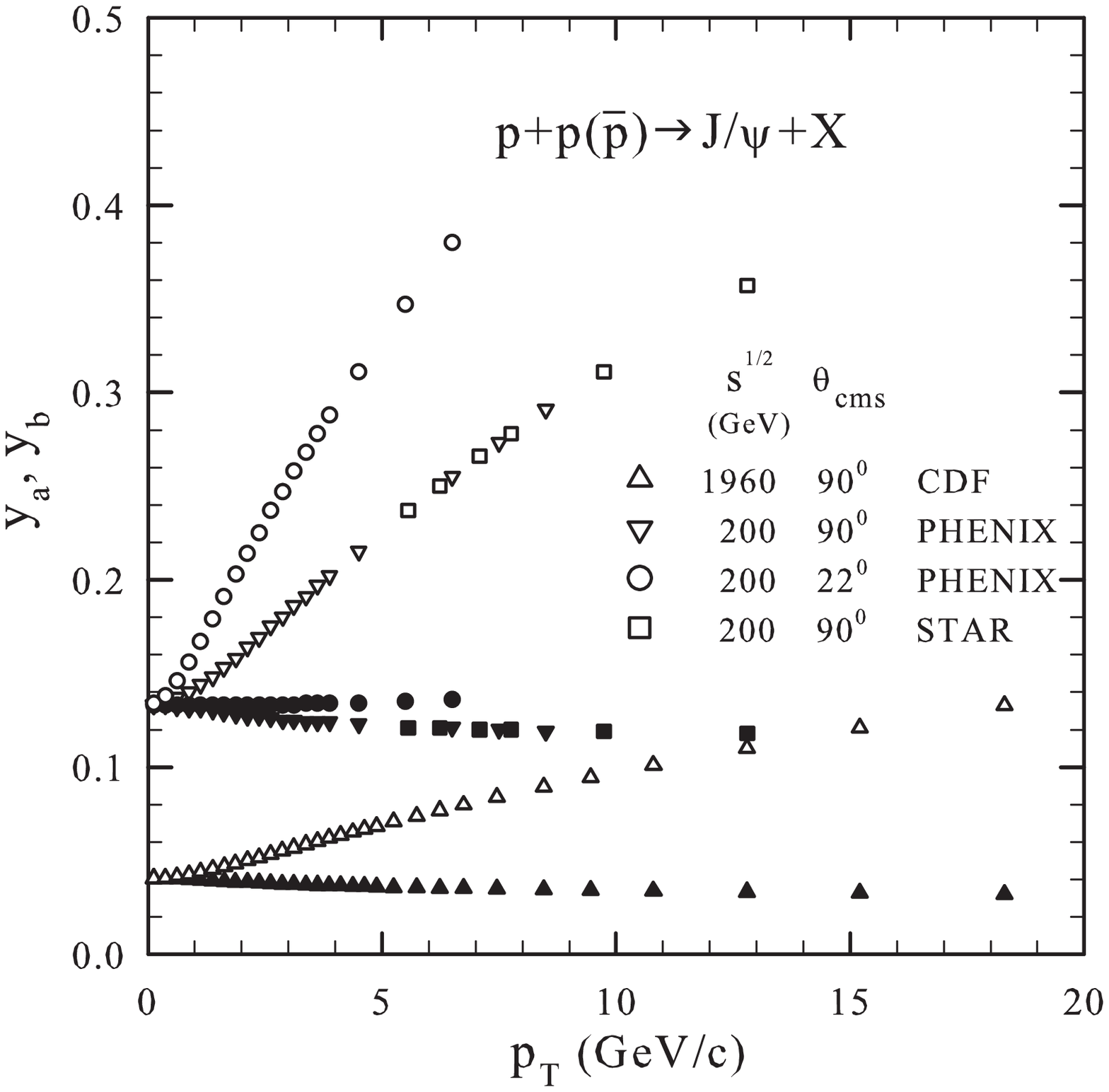}{}} \hspace*{2cm}
\parbox{6cm}{\epsfxsize=4.6cm\epsfysize=4.6cm\epsfbox[85 95 390 400]
{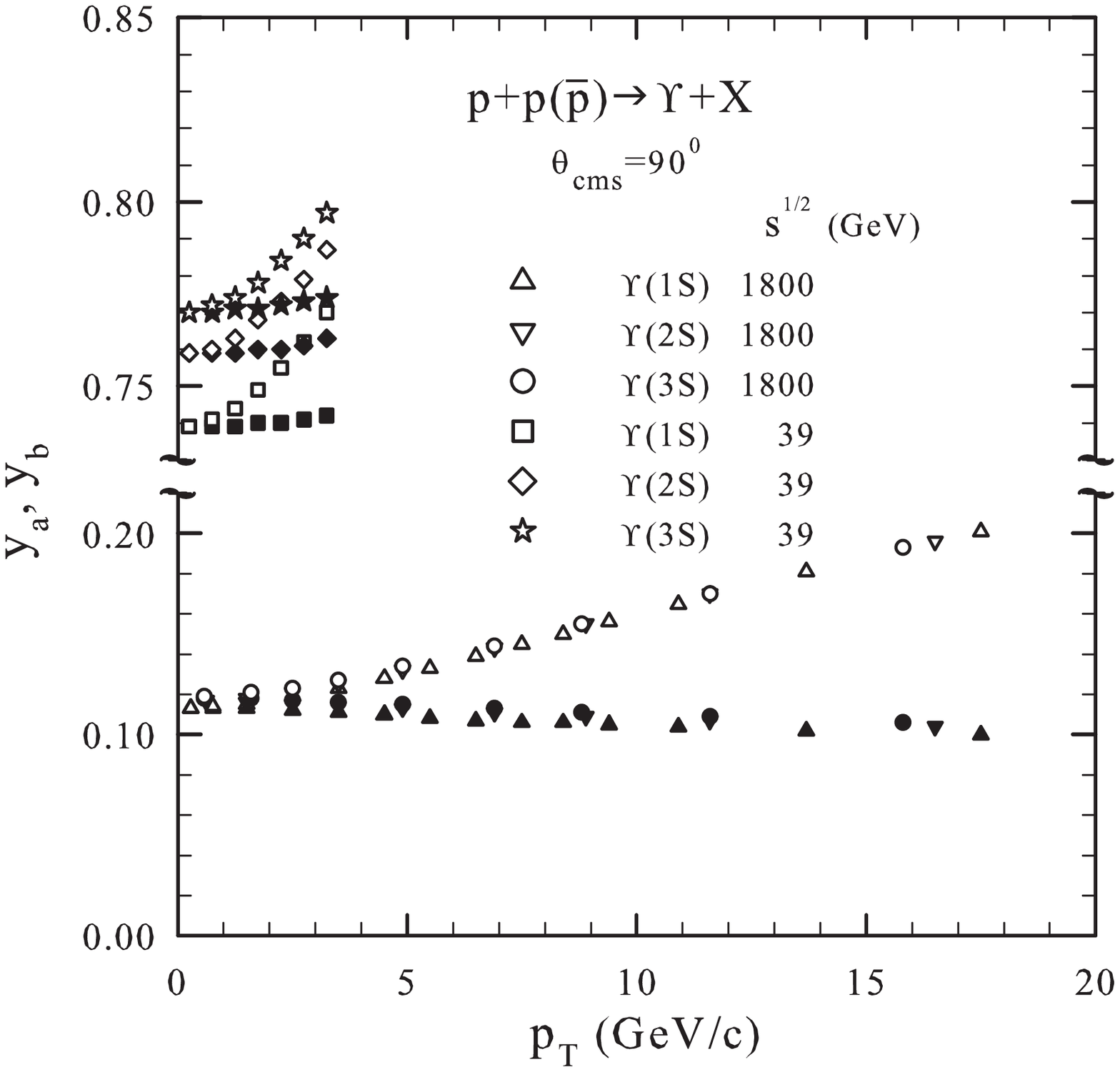}{}} \vskip -1.5cm
\end{center}
\vskip 0.4cm
\hspace*{36mm} (a) \hspace*{7.3cm} (b) \vspace*{8pt}\\
{\bf FIG. 7.}
The dependence of the fractions $y_a$ and $y_b$
on the transverse momentum for
$J/\psi$ (a) and $\Upsilon $ (b) mesons produced in $pp$- and
$\bar pp$-collisions at $\sqrt s = $~200, 1960~GeV and 39, 1800~GeV, respectively.
\vskip 0.5cm

A comparison of the momentum fractions $y_a$ and $y_b$ for the charmonium production
in $pp$\cite{Jpsi_STAR} and $p\bar p$\cite{JPSI_B}
collisions at $\sqrt s =200$  and 1960~GeV is shown in Fig.~7(a).
The fractions follow the general trend i.e. a decrease
with the collisions energy, growth of $y_a$, and approximate
constancy of $y_b$ with $p_T$.
The equality of both fractions at low $p_T$ remains preserved independent on $\sqrt s$.
Another feature typical for all types of the inclusive particles is the angular dependence
of the fractions.
The $y_a$ increases significantly with the decreasing angle $\theta_{cms}$.
It means that the energy dissipation from the constituent subprocess is much smaller in
the fragmentation region as compared to the central interaction region.
The change of the fraction $y_b$ with the angle $\theta_{cms}$ is small.
A levelling of $y_b$ with $p_T$ for small angles is observed.

The sensitivity of the transverse momentum dependence of $y_a$ and $y_b$
to the mass of the $\Upsilon$ states ($1S,2S,3S$) is demonstrated in Fig. 7(b).
The symbols correspond to the experimental data\cite{UPS,UPS_FNAL} on cross
sections measured at FNAL.
It turns out that the momentum fractions  change visibly with the respective state
of the bottomium at smaller collision energy.
The heavier mass the larger fractions.
This is connected with different energy losses
which seem to depend on the mass state of the $\Upsilon$-meson at this energy.
At higher energies the momentum fractions become insensitive
to the single states of this particle.
The curves for the $1S$, $2S$, and $3S$ states coincide
each other at $\sqrt s = 1800$~GeV  in the momentum range $p_T = 0.5-17$~GeV/c.

\vskip 2.5cm
{\subsection{Recoil mass $M_X$ versus $p_T$ and $\sqrt s$}

Another characteristic of the constituent interactions is the recoil mass
\begin{equation}
M_X =x_1M_1+x_2M_2+m_2/y_b
\label{eq:r19}
\end{equation}
released in the underlying subprocess (\ref{eq:r2}).
It is defined by the right-hand side of Eq. (\ref{eq:r3}).
The quantity is proportional to the momentum fractions $x_1$ and $x_2$ of the interacting
objects with the masses $M_1$ and $M_2$.
Its relation to the fractions $y_a$ and $y_b$ is given by the simple dependencies
$x_{1,2}=x_{1,2}(y_a,y_b)$ listed in Ref.~\cite{Gen_Z}.

\vskip 2.5cm
\begin{center}
\hspace*{-0.5cm}
\parbox{6cm}{\epsfxsize=4.6cm\epsfysize=4.6cm\epsfbox[65 95 370 400]
{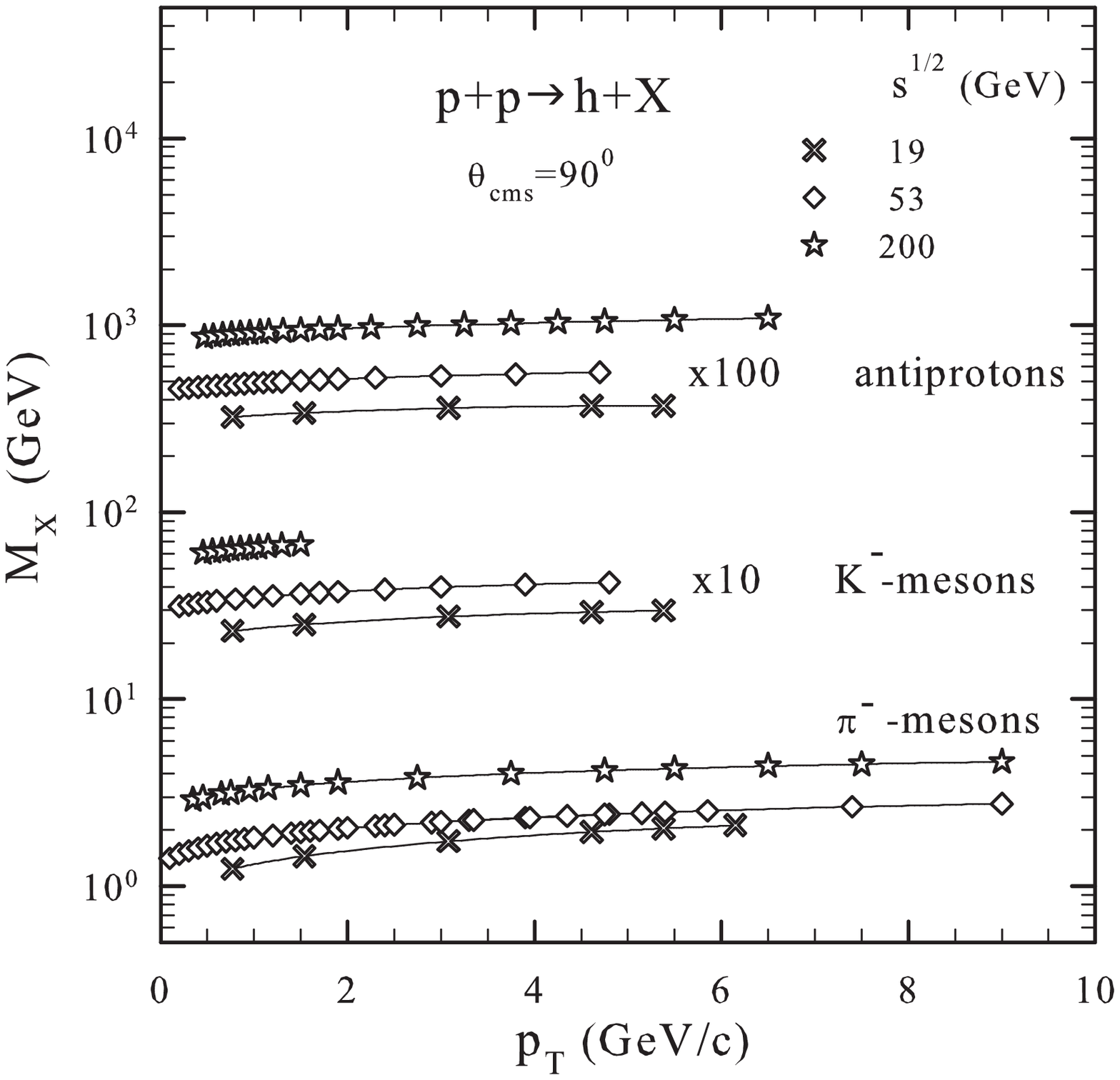}{}} \hspace*{2cm}
\parbox{6cm}{\epsfxsize=4.6cm\epsfysize=4.6cm\epsfbox[85 95 390 400]
{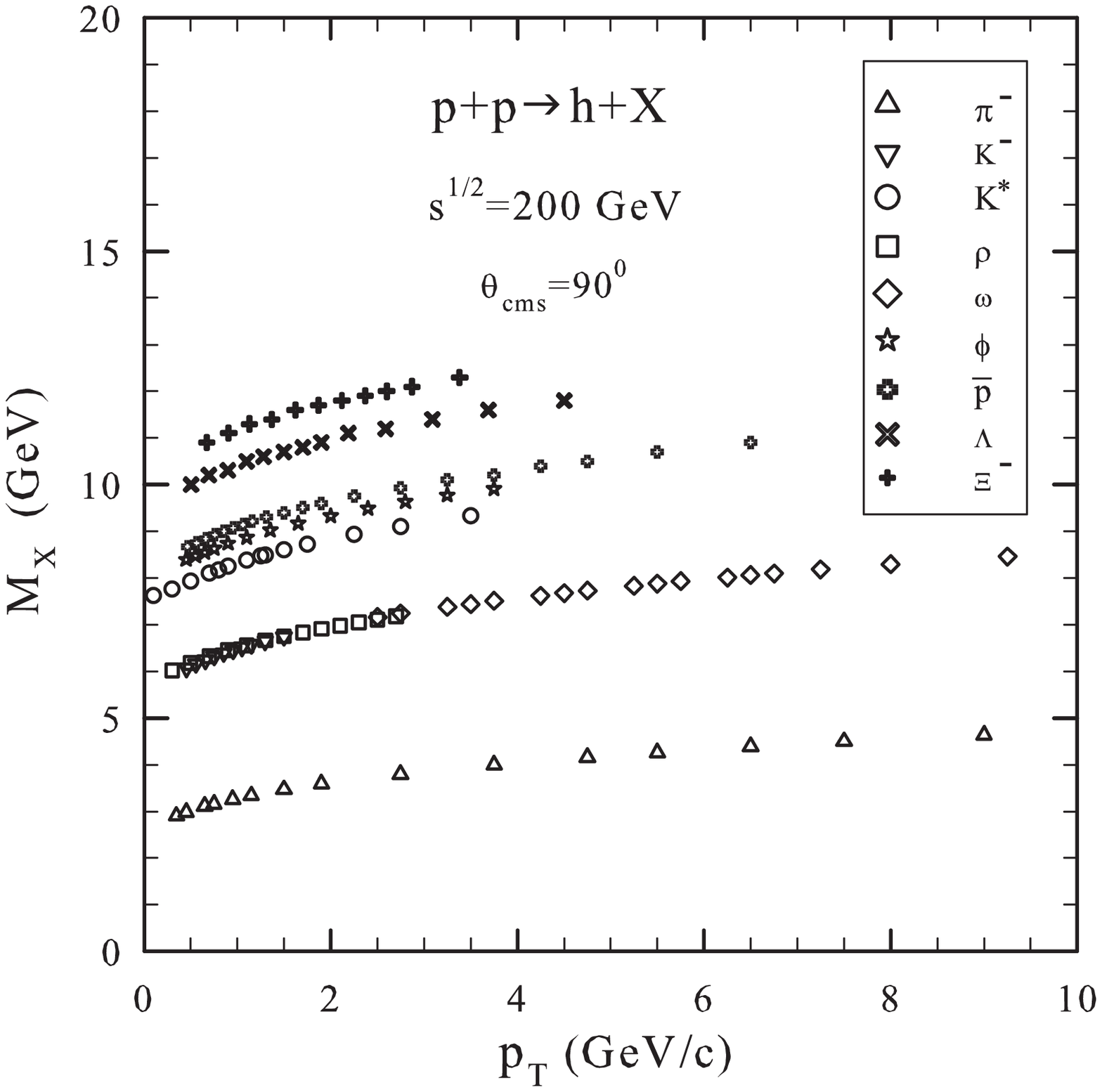}{}} \vskip -1.5cm
\end{center}
\vskip 0.4cm
\hspace*{36mm} (a) \hspace*{7.3cm} (b) \vspace*{8pt}\\
{\bf FIG. 8.}
The recoil mass $M_X$ in the  constituent subprocess underlying
the hadron production in $pp$-collisions at $\sqrt s= 19, 53$, and 200~GeV
in dependence on the transverse momentum $p_T$.
\vskip 0.5cm

The method of determination of $y_a$ and $y_b$ makes them
implicitly dependent on $\delta$ and $\epsilon_F$.
These parameters are obtained from the $z$-scaling approach to analysis
of experimental data on inclusive particle distributions.
In such a way the recoil mass $M_X$ has an internal connection
to the structure of the colliding objects, constituent interactions, and
process of formation of the individual hadrons.
In the very vicinity of the kinematical limit,
the recoil mass is kinematically bounded to approximate the value $M_1+M_2+m_2$.

Figure 8(a) shows the dependence of the recoil mass on the transverse momenta
of the negative pions, kaons, and antiprotons produced in $pp$-collisions
at the energy $\sqrt s = 19, 53$, and $200$~GeV in the central rapidity region.
For the sake of clarity, the values of $M_X$ are presented on a log-scale with
the multiplication factors 10 and 100 for $K^-$ and $\bar p$, respectively.
All curves demonstrate small growth at low $p_T$ followed by a successive flattening.
They reveal a characteristic increase with the collision energy and
mass of the inclusive particle.
The similar dependencies of $M_X$ on $p_T$
for other hadrons measured
at the RHIC energy $\sqrt s=$200~GeV are presented in Fig. 8(b).
The data correspond to the central interaction region.
The dependence of $M_X$ on $p_T$ for $J/\psi, D^0, B$, and $\Upsilon$-meson
production in $p\bar p$-collisions
at the Tevatron energies $\sqrt s=$~1800 and 1960~GeV is presented in Fig. 9(a).

We would like to draw attention to the large recoil mass for the $J/\psi$-meson production.
It is approximately equal to the values of $M_X$ for the $\Upsilon$-mesons
with a small exception at low transverse momenta ($p_T$ below 7~GeV/c).
The mass increases with $p_T$ and reaches the value of $95$~GeV at $p_T = 18$~GeV/c.
The curves for $B$- and $D^0$-mesons lay considerably below.
The extra large values of the recoil mass in the charmonium production
is connected with the abnormal small values
of the corresponding momentum fractions $y_a$ and $y_b$
(see discussion to Fig. 6(b)).
A comparison of the momentum dependence of $M_X$ for $J/\psi$-meson
at $\sqrt s = $~200 and 1960~GeV is shown in Fig. 9(b).
Here one can see the strong sensitivity of the recoil mass
to the collision energy $\sqrt s$.
Its angular dependence at $\sqrt s =200$~GeV is rather weak.
The approximate constancy of $M_X$ at the angle $\theta_{cms}=22^0$
corresponds to the levelling of the fraction $y_b$ with $p_T$ shown in Fig. 7(a).

\vskip 2.5cm
\begin{center}
\hspace*{-0.5cm}
\parbox{6cm}{\epsfxsize=4.6cm\epsfysize=4.6cm\epsfbox[65 95 370 400]
{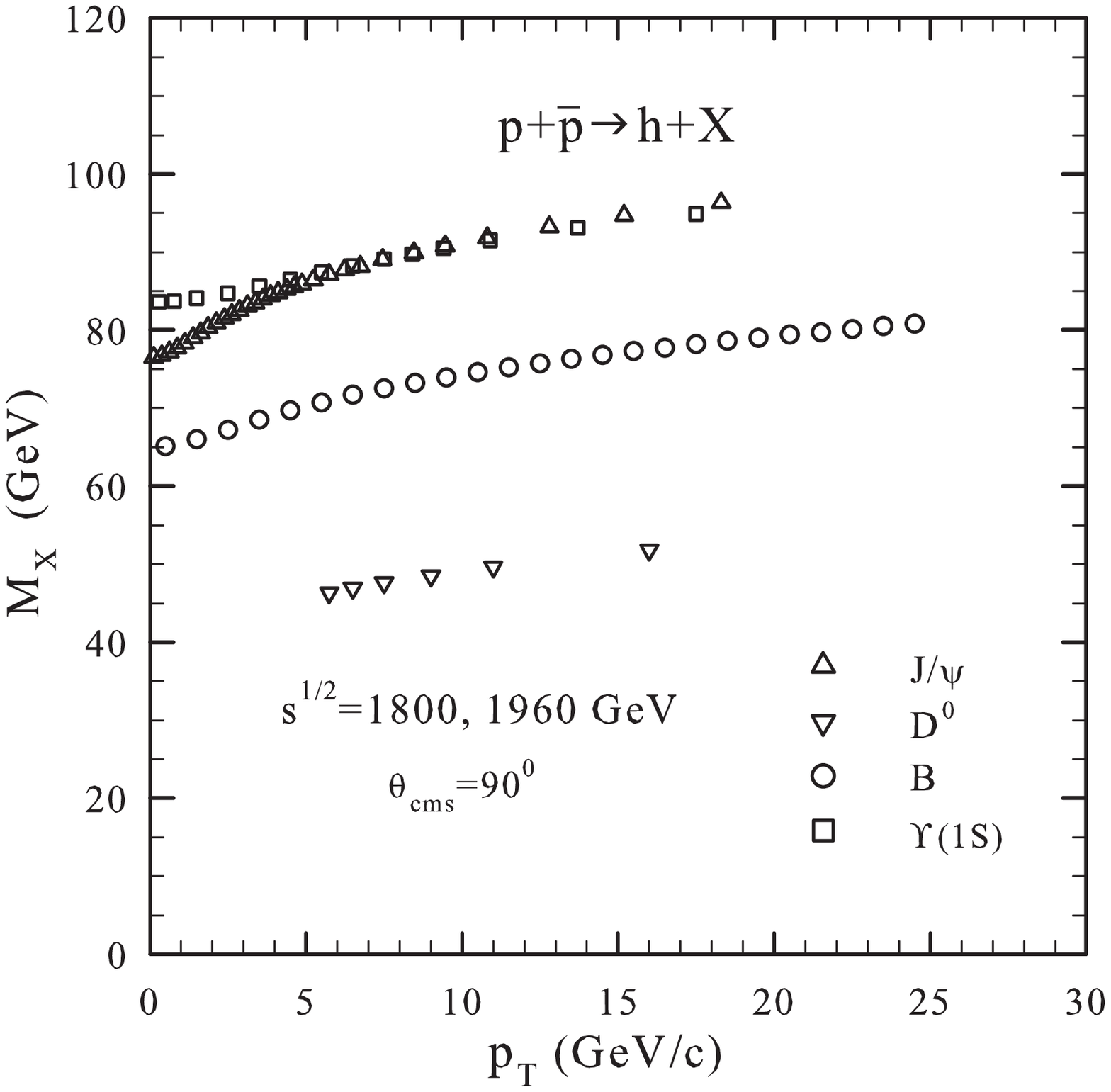}{}} \hspace*{2cm}
\parbox{6cm}{\epsfxsize=4.6cm\epsfysize=4.6cm\epsfbox[85 95 390 400]
{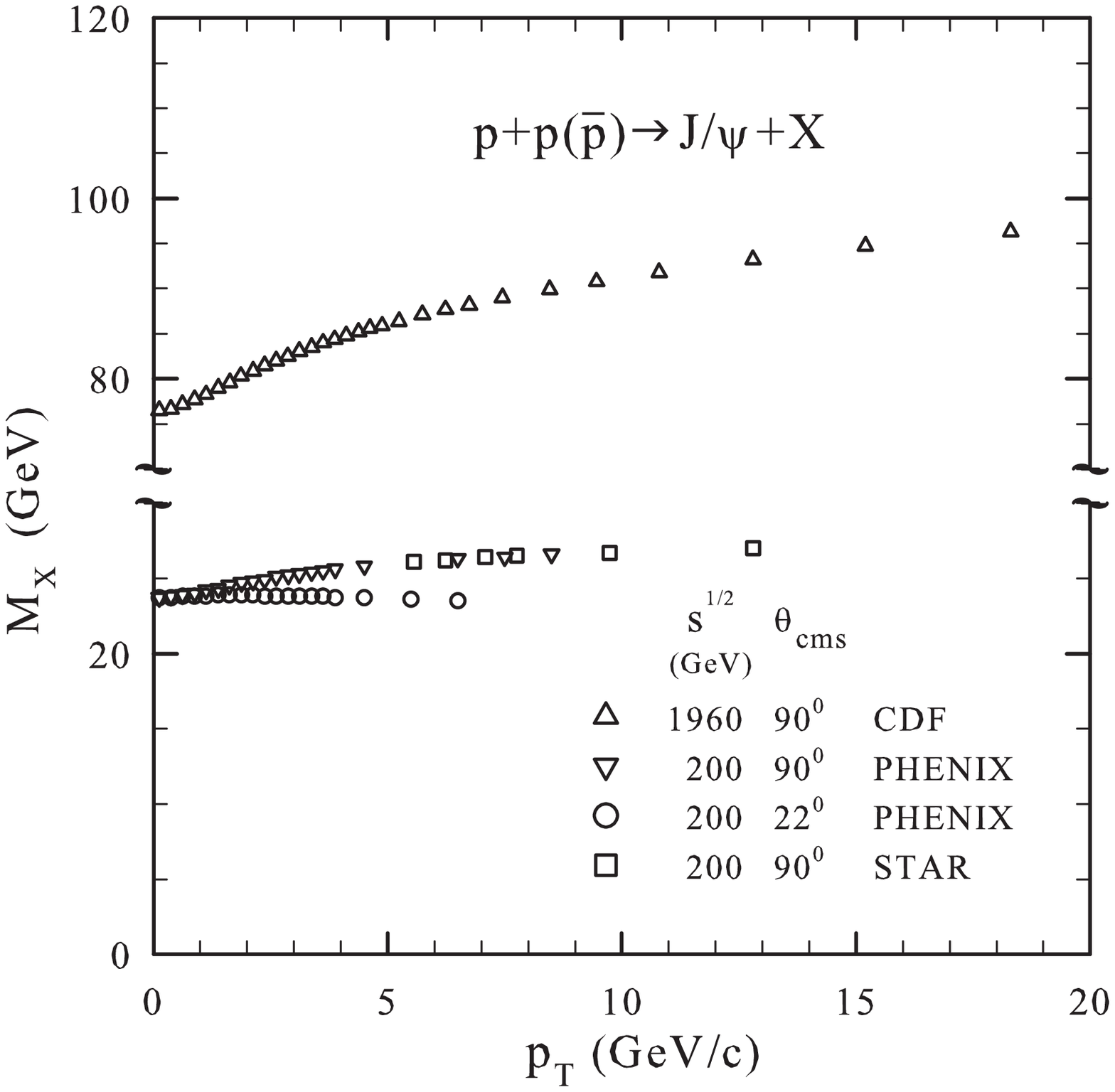}{}} \vskip -1.5cm
\end{center}
\vskip 0.4cm
\hspace*{36mm} (a) \hspace*{7.3cm} (b) \vspace*{8pt}\\
{\bf FIG. 9.}
The recoil mass $M_X$ in the  constituent subprocess underlying production of
(a) $J/\psi$, $D^0$, $B$, $\Upsilon (1S)$ mesons in $p\bar p$-collisions
at $\sqrt s=$~1800, 1960~GeV
and (b) $J/\psi$ mesons in $pp/p\bar p$- collisions at $\sqrt s =$~200, 1960~GeV
in dependence on the transverse momentum $p_T$.
\vskip 0.5cm

\vskip 0.5cm
{\section{Discussion}}

An analysis of ISR, RHIC, and Tevatron experimental data on inclusive
cross sections for different
hadron species performed in $z$-presentation shows the flavor
independence of the shape of the scaling function $\psi(z)$.
This was demonstrated over a wide range of the kinematical variables for
proton-proton and proton-antiproton collisions.
A saturation of $\psi(z)$ was found in the region $z < 10^{-1}$
for hadrons with light and heavy quarks.
The data on $J/\psi$- and $\Upsilon$- meson production
confirm the saturation up to very
small values of $z\simeq 10^{-3}$.
The observed universality is a manifestation of the self-similarity of particle production
at a constituent level.
The dynamical behavior of the interacting system revealed in the distributions
of the inclusive particles is described by the function $\psi(z)$
which demonstrates flattening in the experimentally achievable low $z$-region (Figs. 2 and 3).
The slope $\beta (z)$ of the scaling function diminishes with the decreasing $z$
and becomes zero when the scaling function flattens out.
It was found that for both asymptotic regimes, at low- and high-$z$,
the interacting system is characterized by the equation
\begin{equation}
\frac {d\ln\psi(z)} {d\ln z} = const.
\label{eq:r222}
\end{equation}
The obtained results give strong support that flavor independence of $\psi(z)$
for variety of light and heavy hadrons and its behavior (\ref{eq:r222}) at both asymptotic limits
reflect general trend of hadron production in the high energy proton-proton and
proton-antiproton collisions over a wide range of transverse momenta.

\vskip 0.5cm
{\subsection{Variable $z$ and entropy }}

In this part we discuss analogy between the parameters used in
the $z$-scaling concept  and some thermodynamical
quantities (entropy, specific heat, chemical potential) of a
multiple particle system.
There exists a connection between the variable $z$ and
entropy. The scaling variable is proportional to the ratio
\begin{equation}
z \sim \frac{\sqrt {s_{\bot}}}{{\it W }}
\label{eq:r20}
\end{equation}
of the transverse kinetic energy $\sqrt {s_{\bot}}$
and the maximal value of the function
\begin{equation}
{W(x_1,x_2,y_a,y_b)}=(dN_{ch}/d\eta|_0)^c \  \Omega(x_1,x_2,y_a,y_b)
\label{eq:r21}
\end{equation}
in the space of the momentum fractions.
The quantity $W$ is proportional to all parton and hadron configurations
of the colliding system which can contribute to the production
of the inclusive particle with the momentum $p$.
According to statistical physics, entropy of a system is given by
a number of its statistical states as follows
\begin{equation}
{\it S} = \ln {\it W}.
\label{eq:r22}
\end{equation}
The most likely  microscopic configuration of the system is given
by the maximal value of $\it S$.
In our approach, the configurations comprise all mutually independent
constituent subprocesses (\ref{eq:r2}).
The corresponding subprocesses which can contribute to the production of the
inclusive particle with the 4-momentum $p$ are considered as subject to the
condition (\ref{eq:r3}).
The underlying subprocess, in terms of which the variable $z$ is determined,
is singled out from the corresponding
subprocesses by the principle of maximal entropy $\it S$.

In thermodynamics, entropy for an ideal gas is determined by the formula
\begin{equation}
{\it S} = c_V\ln {T}+ R\ln {V} + {\it S_0},
\label{eq:r23}
\end{equation}
where $c_V$ is a specific heat and $R$ is a universal constant.
The temperature $T$ and the volume $V$ characterize a state of the system.
Using  (\ref{eq:r21}) and (\ref{eq:r22}), we can write the relation
\begin{equation}
{\it S} = c\ \ln {\left[dN_{ch}/d\eta|_0\right]}+
\ln{[(1\!-\!x_1)^{\delta_1}(1\!-\!x_2)^{\delta_2}
(1\!-\!y_a)^{\epsilon_a}(1\!-\!y_b)^{\epsilon_b}]}.
\label{eq:r24}
\end{equation}
Next we exploit analogy between  (\ref{eq:r23}) and (\ref{eq:r24}).
The analogy  is supported by the plausible idea
that interaction of the extended objects like hadrons and nuclei  can be
treated at high energies as a set of independent collisions of their constituents.
Such concept justifies a division of the system into the part
comprising the selected subprocess which underlies production of the inclusive particle
and the rest of the system containing all remaining microscopic configurations
which lead to the produced multiplicity.
The entropy (\ref{eq:r24}) refers to the rest of the system.
The multiplicity density $dN_{ch}/d\eta|_0$ of particles in the central interaction region
characterizes  a "temperature" created in the system \cite{Cleymans}.
Provided that the system is in a local equilibrium, there exists a simple relation
$dN_{ch}/d\eta|_0\sim T^3$ for high temperatures and small chemical potentials.
Using the mentioned analogy, the parameter $c$ plays a role of a "specific heat"
of the produced matter.
The second term in (\ref{eq:r24}) depends on the volume of the rest of the system in the
space of the momentum fractions $\{x_1,x_2,y_a,y_b\}$.
The volume is a product of the complements of the fractions with the exponents which are
generally fractional numbers,
$ {\it V} = l_1^{\delta_1}l_2^{\delta_2}l_3^{\epsilon_a}l_4^{\epsilon_b}$.
This analogy emphasizes once more the interpretation of the parameters
$\delta_1$, $\delta_2$, $\epsilon_a$, and $\epsilon_b$ as fractal dimensions.
In accordance with common arguments, the entropy (\ref{eq:r24}) increases
with the multiplicity density and decreases with the increasing
resolution $\Omega^{-1}$.
In the considered analogy, the principle of a minimal resolution $\Omega^{-1}$
with respect to all subprocesses satisfying the condition
(\ref{eq:r3}) is equivalent to the principle of a maximal entropy $S$
of the rest of the colliding system.

The entropy is determined up to an
arbitrary constant $S_0=\ln W_0$. Dimensional units
entering the definition of the entropy can be included within this
constant. In particular, it allows us to account for a relation between the
dimensionless multiplicity density $dN_{ch}/d\eta|_0$ and the temperature $T$.
This degree of freedom is related to the transformation
\begin{equation}
z\rightarrow  {W_0} \ z, \ \ \ \
\psi \rightarrow {W_0^{-1}} \ \psi.
\label{eq:r25}
\end{equation}
In such a way the scaling variable and the scaling function are
determined up to an arbitrary multiplicative constant. The
constant $W_0$ is related to an absolute number of the microscopic
states of the system. Its value is restricted by the positiveness
of the entropy above some scale characterized by a maximal
resolution $\Omega^{-1}$. For the resolution corresponding to the
fractal limit, $W_0$ is infinity. Thus, the transformation
(\ref{eq:r25}) is connected with a renormalization  of the fractal
measure $z$ in agreement with its physical interpretation (\ref{eq:r20}).
The renormalization represents a shift of the entropy by the corresponding constant,

\begin{equation}
{\it S} = \ln {W} + \ln {W_0}.
\label{eq:r26}
\end{equation}

\vskip 0.5cm
{\subsection{Chemical potential and phase space occupancy}

Let us consider a system where the number of particles can change.
According to the first law of thermodynamics the internal energy $U$ of such system
can be expressed as follows
\begin{equation}
dU= TdS-pdV+\sigma_a dN_a+\sigma_{\bar a} dN_{\bar a}.
\label{eq:r27}
\end{equation}
This shows how $U$ depends on independent variations of the entropy $S$, volume $V$,
and number of particles (antiparticles) $N_a, (N_{\bar a})$.
The parameters $T$ and $p$ are the temperature and pressure of the system.
The potentials $\sigma_{a}$ and $\sigma_{\bar a}$ consist of two parts\cite{Rafelski}
\begin{equation}
\sigma_a \equiv \mu_a+T  \ln (\gamma_a),
\label{eq:r28}
\end{equation}
which are expressed via the chemical potentials
$\mu_{a}=-\mu_{\bar a}$
and the phase space occupancy $\gamma_{a}=\gamma_{\bar a}$ of the system constituents.
First one controls the net number of particles ($N_a-N_{\bar a}$) arising from the particle
and antiparticle difference.
Second term  regulates  the number of particle-antiparticle pairs ($N_a+N_{\bar a}$).
In statistical hadro-chemistry the factors are related to the relative and absolute
chemical equilibrium, respectively \cite{Rafelski}.
The chemical potentials determine the statistical parameter $\Upsilon_a$ known as fugacity
\begin{equation}
\Upsilon_a = \exp (\sigma_a/T)= \gamma_a  \exp (\mu_a/T),
\label{eq:r29}
\end{equation}
which governs the yield of a corresponding particle type.

As shown in Sec. IV, the scaling functions of different hadrons coincide each other
when applying the transformation  (\ref{eq:r17})
with the appropriate values of $\alpha_F$.  The coefficient $\alpha_F$ is
ratio of the constants $W_0$ (\ref{eq:r26}) for single hadrons.
In principle, these properties can be used for estimation of the ratios of the corresponding
phase space occupancy parameters $\gamma_a$.
In Boltzmann approximation, the hadron yields are proportional to the chemical fugacities.
If one assumes that at sufficiently high energy the chemical potentials $\mu_a$ become
negligible on the right-hand side of  (\ref{eq:r28}), then it is possible to write
\begin{equation}
\gamma_1 /\gamma_2 = \alpha_{F_1}/\alpha_{F_2}.
\label{eq:r30}
\end{equation}
The ratio describes the relative yields of hadron pairs with the flavors $F_1$ and $F_2$.
Exploiting the connection between the temperature $T$  and
the multiplicity density $dN/d\eta|_0$  discussed after Eq. (\ref{eq:r24}),
we can rewrite the relation (\ref{eq:r28}) into the form
\begin{equation}
\sigma_{F_1} = \sigma_{F_2} + g(dN_{ch}/d\eta|_0)^{1/3} \ln (\alpha_{F_1}/\alpha_{F_2}).
\label{eq:r31}
\end{equation}
Here $g$ is a constant which does not depend on the particle type.
Knowing the values of $\alpha_F$ for different hadrons (they are quoted in Figs. 2 and 3
relative to $\pi^-$ mesons with $\alpha_{\pi}=1$), one can obtain a hierarchy
of the potentials $\sigma_F$ for the individual hadron species.
A systematic analysis of $\alpha_F$ for  mesons and baryons, its dependence
on the mass, flavor content, and spin requires further, more detailed study.

\vskip 0.5cm
{\subsection{$z$-Scaling and phase transitions}}

The self-similarity of the particle formation at low transverse momenta
is a new specific feature
which can give information on thermodynamical characteristics of
the colliding system and clarify its behavior.
For sufficiently low $p_T$ the scaling variable becomes resolution-independent function of
the multiplicity density $dN_{ch}/d\eta$ and the parameter $c$.
Both quantities characterize  medium produced in the high energy collisions.
The multiplicity density of particles in the central interaction region can be connected with
a "temperature" of the colliding system.
The parameter $c$ is interpreted as a "specific heat" of the produced medium.
Its value $c=0.25$ was found to be the same for different types of the inclusive hadrons.
The estimated error is at the level of 10\%.
We expect that this parameter should change in consequence
of a phase transition in the produced medium.
The increase of the multiplicity (or energy) density
should affect the medium characteristics in that case.
A possible violation of the $z$-scaling in the low-$z$ region connected with a
change of the "specific heat" $c$ might be considered as a manifestation
of such phase transition.
Measurements of the cross sections for hadrons with heavy quarks at higher energies
allow to reach even lower values of $z$ where
the saturation of $\psi(z)$ could in principle change.
We expect that such a change of the observed behavior of the scaling function
could give indications on new physical effects in this region.

There are quantities in thermodynamics which are sensitive to phase transitions.
A discontinuity of specific heat,
compressibility factor and coefficient of thermal expansion
characterizes second order phase transition of the thermodynamic system.
The second derivative of the chemical potential
\begin{equation}
\partial^2 {\sigma} / \partial T^2|_P =-c_P/T
\label{eq:r32}
\end{equation}
is connected with a discontinuity of the specific heat in such a case.
The quantities $c$ in (\ref{eq:r24}) and $\sigma_F$ in (\ref{eq:r31}) signify
a specific heat and chemical potential, respectively.
Their potential dependence on multiplicity density can be studied from suitable experimental data.
A discontinuity of
$c$ associated with a jump in the second derivative of $\sigma_F$,
especially at high $dN_{ch}/d\eta|_0$,
could be considered as a signature of a second order phase transition
in the medium produced in high energy hadron collisions.

It is usually assumed that bulk of the produced matter at low-$p_T$ (low-$z$)
consists of multitude of the strongly interacting constituents.
There is no direct information on the type of the constituents.
Even at high collision energies $\sqrt s$,
the single constituent processes at low transverse momenta are experimentally invisible.
Only indirect signatures can reveal the nature of the constituents.
We suggest to use the analogs of the thermodynamical quantities such as
the specific heat and chemical potentials  which can be useful
in the description of the produced matter.
These quantities can be sensitive to phase transitions in the multiple particle systems.

According to the phase transition theory
and theory of the critical phenomena,
an abrupt change in the specific capacity,
coefficient of thermal expansion and compressibility
with a small change in temperature near to a critical point
is typical for a second-order phase transition \cite{Stanley,Pokr,Ma}.
The second-order phase transitions have a discontinuity in the first
derivative of the entropy
\begin{equation}
c_{V} = \partial {\it S}/\partial \ln (T)|_{V}.
\label{eq:r33}
\end{equation}
In our case, such signature would correspond
to a discontinuity or sharp growth
of the parameter $c$ near some critical point $z_c$.
One can speculate that the value of $c$ should become
dependent on the transverse momentum $p_T$ and
rise sharply enough with the multiplicity density at low $p_T$.
It is usually considered that the increase of the multiplicity density
is connected with a growth of the energy density.
Existence of the internal structure of the constituents
would allow to store the energy into their internal degrees of freedom
which is connected with a rise of the specific heat.
We consider therefore that study of the inclusive cross sections
of hadron production  in the low-$p_T$ region
at high multiplicity densities $dN_{ch}/d\eta$ and
collision energies $\sqrt s$ is most preferable in searching for
signatures of phase transitions in hadron matter.

\vskip 0.5cm
{\section{Conclusions}}

Here we summarize the main results obtained in the paper.
New properties of the $z$-scaling - the flavor independence
and the saturation of $\psi(z)$ at low $z$, were established.
We have studied the spectra of the inclusive particles produced
in proton-proton and proton-antiproton collisions in $z$-presentation.
The experimental data on inclusive  cross sections of different hadrons
obtained at ISR, RHIC, and Tevatron were used in the present analysis.
The data cover a wide range of the collision energies $\sqrt s  = 19-1960$~GeV,
the transverse momenta $p_T = 0.1-10$~GeV/c,
and the production angles $\theta_{cms}=3^0-90^0$.
The variable $z$ depends on the parameters $\delta$, $\epsilon_F$, and $c$.
The parameters $\delta$ and $\epsilon_F$ characterize structure of the colliding (anti)protons
and fragmentation process in the final state, respectively.
Their values are fixed by
the energy, angular, and multiplicity independence of $\psi(z)$
in the high-$p_T$ part of the spectra.
The third parameter $c$ is interpreted as "specific heat" of the produced medium.
The $z$-scaling is consistent with $c=0.25$ and $\delta=0.5$
for all types of the analyzed inclusive hadrons.
The value of $\epsilon_F$ increases with the mass of the produced hadron.
Estimated errors of the determination of the parameters do not exceed 10\%.
It was found that the parameters are independent of kinematical variables.

It was shown that  renormalization of the scaling variable $z$ allows
to reduce the scaling function of different hadrons to a single curve.
The scaling function $\psi(z)$ was established to be flavor independent
in a vide range of $z$.
It means that the shape of $\psi(z)$ is the same for different hadrons in this region.
This includes hadrons with light and heavy quarks  produced
in high energy $pp$- and $p\bar p$-collisions.
A saturation regime of the function $\psi(z)$ was observed for $z<0.1$.
The approximate constancy of $\psi(z)$ was demonstrated up to $z\sim 10^{-3}$ for
charmonia and bottomia.
We conclude that a violation of the asymptotic behavior of the scaling function in the region
of low $z$ could give information on a phase transition of the produced matter.
This should apply especially for the events with high multiplicity densities in both
$pp/p\bar p$- and $AA$-collisions at high energies.

A connection between the ingredients of the variable $z$
allowing thermodynamical interpretation and the thermodynamical quantities
(entropy, temperature, specific heat, chemical potential)
characterizing  multi-particle systems was discussed.
The "specific heat" $c$ was suggested as a parameter useful for characterization
of these systems near a phase transition point.
A microscopic scenario of the constituent interactions in terms of the
momentum fractions was presented.

The universality of description of the
spectra of different hadrons over a wide kinematical region by the scaling function $\psi(z)$
reflects general features (symmetries)
of the underlying physics phenomena over a wide scale range.
Both soft and hard regimes manifest self-similarity of particle production
at a level of hadron constituents.
They are preferable to study collective processes and constituent substructure
at low and high $z$, respectively.

The obtained results may be exploited to search for
and study of new physics phenomena
in particle production in the high energy
proton-proton and proton-antiproton collisions
at U70, RHIC, Tevatron, and LHC.

\vskip 5mm

\acknowledgments  The investigations have been partially supported
by the IRP AVOZ10480505, by the Grant Agency of the Czech Republic
under the contract No. 202/07/0079 and by the special program of
the Ministry of Science and Education of the Russian Federation,
grant RNP.2.1.1.5409.


\end{document}